\documentclass[%
 aip,
 jmp,
 amsmath,amssymb,
 reprint,%
]{revtex4-2}

\usepackage{graphicx}
\usepackage{dcolumn}
\usepackage{bm}

\usepackage{relsize}
\usepackage{graphicx}
\usepackage{dcolumn}
\usepackage{bm}
\usepackage{graphics}
\usepackage{amsmath}
\usepackage{amsfonts}
\usepackage{amssymb}
\usepackage{graphicx}
\usepackage{babel}
\usepackage{mathrsfs}
\usepackage{subfigure}
\usepackage{xcolor}
\usepackage{physics}
\usepackage{bbold}
\usepackage[colorlinks=true, allcolors=blue]{hyperref}
\allowdisplaybreaks

\usepackage[utf8]{inputenc}
\usepackage[T1]{fontenc}
\usepackage{mathptmx}
\usepackage{etoolbox}

\makeatletter
\def\@email#1#2{%
 \endgroup
 \patchcmd{\titleblock@produce}
  {\frontmatter@RRAPformat}
  {\frontmatter@RRAPformat{\produce@RRAP{*#1\href{mailto:#2}{#2}}}\frontmatter@RRAPformat}
  {}{}
}%
\makeatother
\begin{document}


\title[Quantized relativistic TOA-operators for spin-0 particles and the quantum tunneling time problem]{Quantized relativistic time-of-arrival operators for spin-0 particles and the quantum tunneling time problem}
\author{P.C.M. Flores}
\email{pmflores2@up.edu.ph, eagalapon@up.edu.ph}
\altaffiliation[Currently at ]{Max-Born-Institute, Max-Born-Str. 2A, 12489 Berlin, Germany}
\author{E.A. Galapon}%
\affiliation{ 
Theoretical Physics Group, National Institute of Physics, University of the Philippines Diliman, 1101 Quezon City, Philippines
}%

 

\date{\today}

\begin{abstract}
We provide a full account of our recent report [\href{https://iopscience.iop.org/article/10.1209/0295-5075/acad9a/meta}{EPL, \textbf{141} (2023) 10001}] which constructed a quantized relativistic time-of-arrival operator for spin-0 particles using a modified Weyl-ordering rule to calculate the traversal time across a square barrier. It was shown that the tunneling time of a relativistic spin-0 particle is instantaneous under the condition that the barrier height $V_o$ is less than the rest mass energy. This implies that instantaneous tunneling is an inherent quantum effect in the context of arrival times.   
\end{abstract}

\maketitle


\section{Introduction}
\label{sec:intro}

Tunneling is one of the most well-known quantum effects and has been a long standing important subject of quantum mechanics. The simplest tunneling phenomenon is demonstrated by a square potential barrier wherein the Schr\"{o}dinger equation predicts a non-zero probability that a particle initially on the far left of the barrier is transmitted to the far right even if its energy is less than the barrier height. However, tunneling becomes problematic when one associates the time it takes a wavepacket to traverse the classically forbidden region \cite{maccoll1932note,hartman1962tunneling} because it is compounded with the quantum time problem (QTP), and superluminality. Standard quantum mechanics only treats time as a parameter, as such, the quantum tunneling time problem may be ill-defined because there is no canonical formalism in standard quantum mechanics to answer questions regarding time durations \cite{hauge1989tunneling,landauer1994barrier}. Moreover, a dynamical treatment of time, e.g., a time operator, has been met with pessimism because of Pauli's no-go theorem \cite{pauli1933handbuch} on the existence of a time operator. This has led to several definitions of tunneling time using a parametric approach, e.g. Wigner phase time \cite{wigner1955lower}, B\"{u}ttiker-Landauer time \cite{buttiker1982traversal}, Larmor time \cite{baz1966lifetime,rybachenko1967time,buttiker1983larmor}, Pollak-Miller time \cite{pollak1984new}, dwell time \cite{smith1960lifetime}, among many others  \cite{sokolovski1987traversal,yamada2004unified,PhysRevLett.108.170402,de2002time,winful2006tunneling,imafuku1997effects,brouard1994systematic,jaworski1988time,leavens1989dwell,hauge1987transmission}. However, one of us has shown that Pauli's no-go theorem does not hold in the single Hilbert space formulation of quantum mechanics \cite{galapon2002pauli} and constructed a corresponding barrier traversal time operator to calculate the tunneling time \cite{PhysRevLett.108.170402}. By doing so, tunneling time was treated as a dynamical observable which addresses any contentions on tunneling time being an ill-defined problem.

There are still debates on the the validity of the various proposals and its corresponding physical meaning when it predicts apparent superluminal velocities \cite{winful2003nature}. Several experiments \cite{eckleAttosecondAngularStreaking2008,doi:10.1126/science.1163439,pfeifferAttoclockRevealsNatural2012,pfeiffer2013recent,sainadh2019attosecond,torlina2015interpreting,landsman2014ultrafast,camus2017experimental,ramos2020measurement} to measure the tunneling time have confirmed the superluminal behavior of a tunneling particle but there is no consensus on whether the particle is transmitted instantaneously or if it spends a finite time inside the barrier. Moreover, the relation between these various proposed tunneling times is still unclear but it has recently been argued that these tunneling times can be classified into two distinct categories \cite{PhysRevLett.127.133001}, i.e., arrival time and interaction time. The former is concerned with the appearance of the tunneled particle at the far side of the barrier while the latter determines the time duration spent inside the barrier. Tunneling time as an ``arrival time'' is demonstrated by attoclock experiments while ``interaction time'' by Larmor clock  experiments \cite{PhysRevLett.127.133001}. Some attoclock experiments have reported instantaneous \cite{eckleAttosecondAngularStreaking2008,doi:10.1126/science.1163439,pfeifferAttoclockRevealsNatural2012,pfeiffer2013recent,sainadh2019attosecond,torlina2015interpreting} tunneling  while others reported finite tunneling times \cite{landsman2014ultrafast,camus2017experimental}. Moreover, a recent  Larmor clock experiment has also reported finite tunneling time \cite{ramos2020measurement}. Now, whether tunneling is instantaneous or not, the crux of the problem is that both results imply that the particle exhibits superluminal behavior below the barrier. This now raises the question on whether the superluminality is a consequence of using non-relativistic quantum mechanics, i.e., could there a fundamental difference if one uses a relativistic theory? 

There have been several studies to extend the analysis of tunneling times to the relativistic case in order to adequately address the superluminal behavior \cite{de2007dirac,de2013study,PhysRevA.67.012110,krekora2001effects}. It was shown by de Leo and Rotelli \cite{de2007dirac}, then separately again by de Leo \cite{de2013study} whom used the phase time via the Dirac equation in a step potential to show  that superluminal tunneling times is still present.  Petrillo and Janner \cite{PhysRevA.67.012110} obtained similar results for a square barrier via the Dirac equation. Krekora, Su, and Grobe \cite{krekora2001effects} also used the Dirac equation for various potential barriers of the form $V(x)=V_o e^{-(2x/w)^n}$ with an effective width $w$, and defined an ``instantaneous tunneling speed'' to show superluminal tunneling under the condition that the barrier height $V_o$ is less than twice the rest mass energy. This apparent superluminal behavior despite the relativistic treatment implies that the superluminal behavior is an inherent quantum effect.  

In this paper, we give a full account of our recent report \cite{flores2022letter} which proposed a formalism on the construction of quantized relativistic TOA-operators for spin-0 particles in the presence of an interaction potential. This was then used to construct a corresponding barrier traversal time operator. By doing so, the formalism can simultaneously addresses the compounding problems of superluminality and the QTP in tunneling times. Now, it is well-known that relativistic quantum mechanics is not a well-defined one-particle theory since relativistic effects can lead to spontaneous pair-creation and annihilation which might render the concept of TOA meaningless, i.e., we are not sure if the particle that tunneled and arrived is the same particle we initially had. To address this, we will impose the condition that the barrier height is less than the rest mass energy.

The rest of the paper is structured as follows. In Sec. \ref{sec:RevNonrel} we review the construction of quantized non-relativistic TOA-operators in coordinate representation using Weyl, Born-Jordan, and simple symmetric ordering \cite{galapon2018quantizations} which will then be modified to construct the corresponding relativistic counterpart for spin-0 particles in Sec. \ref{sec:PosSpaceRep}. The barrier traversal time operator is then constructed in Sec. \ref{sec:timeopr} and will be shown to reduce to the correct classical limit as $\hbar\rightarrow0$ in Sec. \ref{sec:ClassLim}. Next, we establish the expected barrier traversal time and show that tunneling is instantaneous in Sec. \ref{sec:expec}, regardless of the ordering rule used. A single Gaussian wavepacket is then used as an example in Sec. \ref{sec:Gaussian}. Last, we conclude in Sec. \ref{sec:conc}.

\section{Review of quantized non-relativistic TOA-operators}
\label{sec:RevNonrel}

The rigorous mathematical framework of quantum mechanics was developed by von Neumann using the Hilbert space $\mathcal{H}$ as its underlying linear topological space wherein physical observables are generally identified with maximally symmetric densely defined operators $\mathsf{\hat{A}}$ in $\mathcal{H}$ while physical states are represented by the set of unit rays $\ket{\psi}$ in $\mathcal{H}$. The eigenvalues of these operators then represent the possible measurement outcomes of the corresponding observable and its spectrum may be discrete, continuous, or a combination of both. However, operators in quantum mechanics are generally unbounded with a continuous spectrum corresponding to non-normalizable eigenfunctions, e.g. the position and momentum operator whose eigenfunctions are the Dirac-delta function $\delta(q-q_o)$ and the plane wave $\exp(ipq/\hbar)/\sqrt{2\pi\hbar}$, respectively. 

In order to deal with these non-square integrable functions that are outside the Hilbert space, one can use Dirac's bra-ket notation which is made mathematically rigorous by using the rigged Hilbert space (RHS) that utilizes the theory of distributions \cite{bohm1974rigged,de2002rigged,de2002rigged2,de2003rigged,galapon2018quantizations}. In our case, we choose the fundamental space of our RHS to be the space of infinitely continuously differentiable complex valued functions with compact supports $\Phi$ such that the RHS is $\Phi\subset L^2(\mathbb{R}) \subset \Phi^\times$, where $\Phi^\times$ is the space of all continuous linear functionals on $\Phi$. The standard Hilbert space formulation of quantum mechanics is recovered by taking the closures on $\Phi$ with respect to the metric of $L^2(\mathbb{R})$. 

In coordinate representation, a quantum observable $\mathsf{\hat{A}}$ is a mapping from $\Phi$ to $\Phi^\times$, and is given by the formal integral operator 
\begin{equation}
	(\mathsf{\hat{A}}\varphi)(q) = \int_{-\infty}^\infty dq' \mel{q}{\mathsf{\hat{A}}}{q'} \varphi(q')
	\label{eq:genintop}
\end{equation}
where the kernel satisfies $ \mel{q}{\mathsf{\hat{A}}}{q'}= \mel{q'}{\mathsf{\hat{A}}}{q}^*$, to ensure Hermiticity such that the eigenvalues of Eq. \eqref{eq:genintop} are real-valued. The integral Eq. \eqref{eq:genintop} is interpreted in the distributional sense, i.e. it is a functional on $\Phi$ wherein the kernel $\mel{q}{\mathsf{\hat{A}}}{q'}$ is a distribution. As an example, the position and momentum operators are now given as 
\begin{align}
	(\mathsf{\hat{q}}\varphi)(q) =& \int_{-\infty}^\infty dq' \delta(q-q') \varphi(q') = q\varphi(q) \\ 
	(\mathsf{\hat{p}}\varphi)(q) =& \int_{-\infty}^\infty dq' i\hbar \dfrac{d \delta(q-q')}{dq'} \varphi(q') = -i\hbar \dfrac{d \varphi(q)}{dq}.
\end{align}

There is still no consensus on how TOA-operators in the presence of an interaction potetial are constructed \cite{leon2000time,peres2006quantum,galapon2018quantizations}. One possible method is by canonical quantization but it has been deemed not meaningful because the classical TOA can be multiple and/or complex-valued. Moreover, canonical quantization suffers from ordering ambiguities, obstruction to quantization \cite{gotay2000obstruction,groenewold1946principles}, and circularity when imposing the correspondence principle \cite{Galapon2001,galapon2004shouldn}. To overcome these problems, the method of ``supraquantization'' was proposed for non-relativistic TOA-operators \cite{galapon2004shouldn}, i.e., constructing TOA-operators from first principles of quantum mechanics. It turns out that for linear systems, $V(q)=\alpha q^2 + \beta q + \gamma$, the ``supraquantized'' TOA-operator is equal to the canonically quantized TOA-operator using Weyl-ordering. Meanwhile, the ``supraquantized'' TOA-operator for non-linear systems can be expressed as a perturbation series wherein the leading term is the Weyl-ordered TOA-operator \cite{galapon2004shouldn,galapon2018quantizations,pablico2022quantum}. This relation shows that canonical quantization is sufficient as a leading order approximation of the canonical TOA-operator.

Now, the non-relativistic TOA-operators constructed by Galapon and Magadan \cite{galapon2018quantizations} quantized the corresponding classical non-relativistic TOA
\begin{equation}
	t_x(q,p) = -\text{sgn}(p) \sqrt{\dfrac{\mu_o}{2}} \int_x^q dq' \left[ \dfrac{p^2}{2\mu_o} + V(q)-V(q') \right]^{-1/2}
	\label{eq:nonreltoa}
\end{equation}
in coordinate representation. The function $\text{sgn}(p)$ is the sign of the initial momentum $p$ which accounts for the particles moving from the left or right. Meanwhile, $x$ is the arrival point and $\mu_o$ is the rest mass of the particle. The expression Eq. \eqref{eq:nonreltoa} was obtained by treating energy as a constant of motion and inverting the corresponding Hamilton equation of motion. To quantize Eq. \eqref{eq:nonreltoa}, it has been argued \cite{galapon2018quantizations} that objections can be addressed on physical grounds. First, the TOA of a quantum particle is always real-valued because it can tunnel to the classically forbidden region. Second, it is only meaningful to quantize the first TOA because the wavefunction will collapse after a detector registers the TOA of the quantum particle. The quantization of Eq. \eqref{eq:nonreltoa} was done by first expanding  around the free TOA \cite{galapon2018quantizations}, i.e.,
\begin{equation}
	t_0(q,p)=-\sum_{j=0}^\infty (-1)^j \mu_o^{j+1} \dfrac{(2j-1)!!}{j! p^{2j+1}}\int_o^q dq'   \left( V(q) - V(q')\right)^j.  
\end{equation} 
It is then assumed that the potential is analytic at the origin wherein it admits the expansion $V(q)=\sum_{n=0}\nu_n q^n$ such that
\begin{align}
	\int_o^q dq'   \left( V(q) - V(q')\right)^j = \sum_{n=1}^\infty a_n^{(j)} q^n.
	\label{eq:potential}
\end{align}
This then yields the local time of arrival (LTOA)
\begin{equation}
	t_0(q,p)=-\sum_{j=0}^\infty (-1)^j \mu_o^{j+1} \dfrac{(2j-1)!!}{j!}  \sum_{n=1}^{\infty} a_n^{(j)}  \dfrac{q^n}{p^{2j+1}},
	\label{eq:LTOA}
\end{equation}
which is now amenable to quantization because it is single and real-valued in its region of convergence in the phase space. Now, the LTOA converges absolutely and uniformly only in some local neighborhood $\omega = \omega_q \times \omega_p$ determined by $\abs{V(q)-V(q')}<p^2/2\mu_o$ for $p\neq0$ and continuous $V(q)$, and will diverge outside this region which signifies that the particle has not classically arrived at $q=0$, i.e. the classically forbidden region. However, the classical TOA Eq. \eqref{eq:nonreltoa} holds in the region $\Omega=\Omega_q\times\Omega_p$ where $\omega\subset\Omega$. This means that Eq. \eqref{eq:nonreltoa} is the analytic continuation of the LTOA in the region $\Omega \backslash \omega$ \cite{pablico2022quantum}.

The monomials $q^n p^{-m}$ were then quantized by generalizing the Bender-Dunne basis operators \cite{bender1989exact,bender1989integration}, 
\begin{equation}
	\mathsf{\hat{t}_{-m,n}} = \dfrac{\sum_{k=0}^n \beta_k^{(n)} \mathsf{\hat{q}}^k\mathsf{\hat{p}}^{-m}\mathsf{\hat{q}}^{n-k}}{\sum_{k=0}^n \beta_k^{(n)}}, 
	\label{eq:BDbasis}
\end{equation} 
where, the coefficients satisfy the condition $\beta_k^{(n)}=\beta_{n-k}^{(n)*}$ to ensure Hermiticity. Now, the most well-studied \cite{domingo2015generalized,de2016born,cohen2012weyl,Gosson2016,Gosson2016a,Gosson2011} ordering rules are Weyl, Born-Jordan, and simple-symmetric with each having its own advantage . Specifically, Weyl ordering preserves the covariant property of Hamiltonian dynamics with respect to linear canonical transforms \cite{Gosson2016,de2006symplectic} while Born-Jordan preserves the equivalence of the the Schr\"{o}dinger and Heisenberg formulation of quantum mechanics \cite{Gosson2016,de2013born,cohen1966generalized}. On the other hand, simple-symmetric ordering just provides the easiest possible ordering by using the ``average rule'' \cite{domingo2015generalized,shewell1959formation}. These ordering rules are imposed on the basis operators $\mathsf{\hat{t}_{-m,n}}$ by choosing the coefficients
\begin{align}
	\beta_k^{(n)} = 
	\begin{cases}
		\dfrac{n!}{k!(n-k)!} \quad &, \quad \text{Weyl} \\
		1 \quad &, \quad \text{Born-Jordan} \\
		\delta_{k,0} + \delta_{k,n} \quad &, \quad \text{simple-symmetric}. 
	\end{cases}
	\label{eq:coeff}
\end{align}

It easily follows that in coordinate representation, the non-relativistic TOA-operator admits the expansion  
\begin{equation}
	(\mathsf{\hat{T}_0}\varphi)(q) = -\int_{-\infty}^\infty dq' \sum_{j=0}^\infty (-1)^j \mu_o^{j+1} \dfrac{(2j-1)!!}{j!}  \sum_{n=1}^{\infty} a_n^{(j)}  \mel{q}{\mathsf{\hat{t}_{-2j-1,n}}}{q'} \varphi(q').
	\label{eq:genTOAoprExpand}
\end{equation}
wherein 
\begin{align}
	\mel{q}{\mathsf{\hat{t}_{-m,n}}}{q'} = \dfrac{i(-1)^{\frac{1}{2}(m-1)}}{2\hbar^m(m-1)!}P_n(q|q')(q-q')^{m-1}\text{sgn}(q-q'), \quad m=1,2,\dots
	\label{eq:BDkernel}
\end{align}
\begin{align}
	P_n(q|q') = 
	\begin{cases}
		\left( \dfrac{q+q'}{2}\right)^n \quad&, \quad \text{Weyl} \\ \\
		\dfrac{1}{n+1} \left( \dfrac{q^{n+1} - q'^{n+1}}{q-q'} \right) \quad&, \quad \text{Born-Jordan} \\ \\
		\dfrac{q^n+q'^n}{2} \quad&, \quad \text{simple-symmetric}.
	\end{cases}
	\label{eq:basisFactors}
\end{align}
The summation over $n$ in Eq. \eqref{eq:genTOAoprExpand} is then evaluated using the following identities
\begin{align}
	\sum_{n=1}^{\infty} a_n^{(j)} P_n(q|q') = 
	\begin{cases}
		\mathlarger\int_0^{(q+q')/2} ds \left[ V\left(\dfrac{q+q'}{2}\right) - V(s)\right]^j  \quad&, \quad \text{Weyl} \\ \\ 
		\mathlarger\int_0^qdu \mathlarger\int_0^s (V(s)-V(u))^j - \mathlarger\int_0^{q'}du \mathlarger\int_0^s (V(s)-V(u))^j \quad&, \quad \text{Born-Jordan} \\ \\
		\dfrac{1}{2} \mathlarger\int_0^q ds (V(q)-V(s))^j + \dfrac{1}{2} \mathlarger\int_0^{q'} ds (V(q)-V(s))^j \quad&, \quad \text{simple-symmetric},
		\label{eq:formfactor}
	\end{cases}
\end{align}
which follows from the assumed analyticity of the potential at the origin Eq. \eqref{eq:potential}. The resulting expression is further evaluated by taking the summation over $j$. 

Performing these operations yield the non-relativistic TOA-operators of the form
\begin{equation}
	(\mathsf{\hat{T}_0}\varphi)(q) = \int_{-\infty}^\infty dq' \dfrac{\mu_o}{i\hbar} T_0(q,q') \text{sgn}(q-q') \varphi(q'),
	\label{eq:genTOAopr}
\end{equation}
where $T(q,q')$ is referred to as the time kernel factor (TKF) which depends on the ordering rule used, i.e.,
\begin{align}
	T_0^{(W)}(q,q') =& \dfrac{1}{2} \int_0^{\frac{q+q'}{2}} ds {_0}F_1\left[;1;\dfrac{\mu_o}{2\hbar^2}(q-q')^2 \left\{ V\left(\frac{q+q'}{2}\right) - V(s)\right\}\right] \label{eq:nonrelWeyl}\\
	\nonumber \\
	T_0^{(BJ)}(q,q') =& \dfrac{1}{2(q-q')} \int_0^q ds \int_0^s du {_0}F_1\left[;1;\dfrac{\mu_o}{2\hbar^2}(q-q')^2 \left\{ V(s) - V(u)\right\}\right] \nonumber \\
	&-\dfrac{1}{2(q-q')} \int_0^{q'} ds \int_0^s du {_0}F_1\left[;1;\dfrac{\mu_o}{2\hbar^2}(q-q')^2 \left\{ V(s) - V(u)\right\}\right] \label{eq:nonrelBJ} \\
	\nonumber \\
	T_0^{(SS)}(q,q')=& \dfrac{1}{4} \int_0^q ds  {_0}F_1\left[;1;\dfrac{\mu_o}{2\hbar^2}(q-q')^2 \left\{ V(q) - V(s)\right\}\right] \nonumber \\
	&+ \dfrac{1}{4} \int_0^{q'} ds  {_0}F_1\left[;1;\dfrac{\mu_o}{2\hbar^2}(q-q')^2 \left\{ V(q') - V(s)\right\}\right] \label{eq:nonrelSS}
\end{align}
where ${_0}F_1(;a;z)$ is a specific hypergeometric function. The superscripts ``W'', ``BJ'', and ``SS'' refer to the Weyl, Born-Jordan, and simple symmetric ordering, respectively.

\section{Non-analytic quantization of the relativistic LTOA in coordinate representation}
\label{sec:PosSpaceRep}

Supposing that the relation between the non-relativistic ``supraquantized'' and quantized TOA-operators also holds for the relativistic case, it should be enough for now to consider the simplest approach by developing the method of canonical quantization of relativistic TOA-operators, and leave the method of ``supraquantization'' open for future studies. We follow the steps outlined in Sec. \ref{sec:RevNonrel} to construct the relativistic TOA-operator by quantizing the corresponding ``classical'' relativistic time-of-arrival (CRTOA) obtained from inverting the equation of motion from the Hamiltonian of special relativity \cite{greiner2000relativistic}, i.e.,
\begin{align}
	t_x(q,p) =& -\text{sgn}p\int_x^q \dfrac{dq'}{c} \left(1-\dfrac{\mu_o^2c^4}{(H(q,p)-V(q'))^2}\right)^{-1/2}
	\label{eq:classreltoa0}
\end{align}
wherein 
\begin{equation}
	H(q,p) = \sqrt{p^2c^2+\mu_o^2c^4} + V(q)
\end{equation}
is the total energy of the positive energy solutions generated by the Klein-Gordon equation. Similar to Eq. \eqref{eq:nonreltoa}, the expression Eq. \eqref{eq:classreltoa0} was also obtained by treating energy as a constant of motion and inverting the corresponding Hamilton equation of motion. Without loss of generality, we assume the arrival point to be the origin $x=0$ and impose that the potential is analytic at the origin such that Eq. \eqref{eq:classreltoa0} has the expansion around the relativistic free TOA given by 
\begin{align}
	t_0(q,p)=& - \mu_o \sum_{j=0}^\infty \sum_{k=0}^j \binom{-\frac{1}{2}}{j} \binom{j}{k} \dfrac{(2\mu_o)^j}{(2\mu_o c^2)^{j-k}} \sum_{n=1}^\infty a_n^{(2j-k)} \dfrac{\gamma_p^{k+1}}{p^{2j+1}} q^n
	\label{eq:expand2quant} 
\end{align}
where, $\gamma_p = \sqrt{1 + p^2/\mu_o^2c^2}$. For consistency with Sec. \ref{sec:RevNonrel}, we shall also refer to Eq. \eqref{eq:expand2quant} as the relativistic LTOA since it is also single and real-valued within its region of convergence in the phase space. That is, Eq. \eqref{eq:expand2quant} will only converge absolutely and uniformly in some local neighborhood $\omega=\omega_q\times\omega_p$ determined by $\abs{V(q)-V(q')} > \sqrt{p^2c^2 + \mu_o^2 c^4}-\mu_o c^2$ for $p\neq0$ and continuous $V(q)$. Meanwhile Eq. \eqref{eq:classreltoa0} holds in the region $\Omega=\Omega_q\times\Omega_p$ where $\omega\subset\Omega$ and is the analytic continuation of the relativistic LTOA in the region $\Omega\backslash\omega$. 

\begin{figure}[t!]
	\centering
	\includegraphics[width=0.95\textwidth]{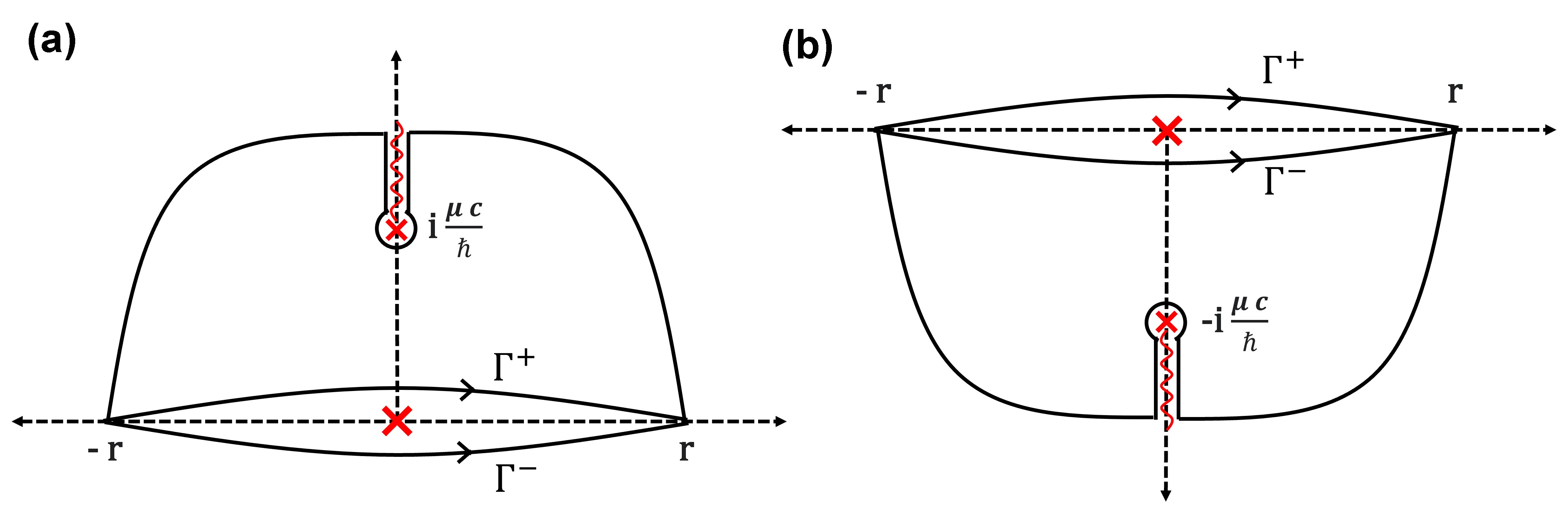}
	\caption{Contours of integration for Eq. \eqref{eq:pkernelGen} for (a) $q-q'>0$ and (b) $q-q'<0$. }
	\label{fig:contour_pos} 
\end{figure}	

The relativistic LTOA Eq. \eqref{eq:expand2quant} is now amenable to quantization by promoting the position and momentum $(q,p)$ into operators $(\mathsf{\hat{q}},\mathsf{\hat{p}})$. There is still no consensus on the existence of a position operator in relativistic quantum mechanics \cite{leon1997time} but the most suitable candidate is the Newton-Wigner position operator \cite{newton1949localized}. In our case, we will only use the non-relativistic position operator $\mathsf{\hat{q}}$ in quantizing Eq. \eqref{eq:expand2quant} which is motivated by Razavi's relativistic free TOA operator \cite{razavy1969quantum,flores2022relativistic}
\begin{equation}
	\mathsf{\hat{T}_{Ra}} = -\dfrac{\mu_o}{2} \left( \mathsf{\hat{q}} \mathsf{\hat{p}^{-1}} \sqrt{1 + \dfrac{\mathsf{\hat{p}^2}}{\mu_o^2 c^2}} + \mathsf{\hat{p}^{-1}} \sqrt{1 + \dfrac{\mathsf{\hat{p}^2}}{\mu_o^2 c^2}} \mathsf{\hat{q}} \right). 
\end{equation} 
To quantize Eq. \eqref{eq:expand2quant}, we extend the Bender-Dunne basis operators \cite{bender1989exact,bender1989integration} to separable classical function $f(q,p)=g(q)^nh(p)^m$, i.e.,
\begin{equation}
	f(q,p) \Rightarrow \mathsf{\hat{f}_{\hat{q},\hat{p}}} = \dfrac{\sum_{k=0}^n \alpha_k^{(n)} \mathsf{\hat{g}_{\hat{q}}}^k\mathsf{\hat{h}_{\hat{p}}}^m\mathsf{\hat{g}_{\hat{q}}}^{n-k}}{\sum_{k=0}^n \alpha_k^{(n)}}.
	\label{eq:BDbasisGen}
\end{equation}
where the coefficients $\alpha_k^{(n)}$ are given by Eq. \eqref{eq:coeff}. This now leads to the quantization 
\begin{align}
	Q\left[ q^n p^{-2j-1} \gamma_p^{k+1} \right] = 
	\begin{cases}
		\dfrac{1}{2^n} \mathlarger\sum_{r=0}^n \binom{n}{r} \mathsf{\hat{q}}^{r} \mathsf{\hat{p}}^{-2j-1} \mathsf{\gamma_{\hat{p}}^{k+1}} \mathsf{\hat{q}}^{n-r} \quad&, \quad \text{Weyl} \\ \\
		\dfrac{1}{n+1} \mathlarger\sum_{r=0}^n \mathsf{\hat{q}}^{r} \mathsf{\hat{p}}^{-2j-1} \mathsf{\gamma_{\hat{p}}^{k+1}} \mathsf{\hat{q}}^{n-r} \quad&, \quad \text{Born-Jordan} \\ \\
		\dfrac{1}{2} \left( \mathsf{\hat{q}}^{n} \mathsf{\hat{p}}^{-2j-1} \mathsf{\gamma_{\hat{p}}^{k+1}} +  \mathsf{\hat{p}}^{-2j-1} \mathsf{\gamma_{\hat{p}}^{k+1}} \mathsf{\hat{q}}^{n} \right) \quad&, \quad \text{simple-symmetric}
	\end{cases}
	\label{eq:NonAnaQuant}
\end{align}

It follows from Eq. \eqref{eq:NonAnaQuant} that in coordinate representation, the quantized relativistic TOA Eq. \eqref{eq:expand2quant} now has the expansion 
\begin{align}
	(\mathsf{\hat{T}_c}\varphi)(q)= - \mu_o \sum_{j=0}^\infty \sum_{k=0}^j & \binom{-\frac{1}{2}}{j} \binom{j}{k} \dfrac{(2\mu_o)^j}{(2\mu_o c^2)^{j-k}} \nonumber \\
	&\times \sum_{n=1}^\infty a_n^{(2j-k)} \int_{-\infty}^{\infty} dq'   P_n^{(Q)}(q|q') \mel{q}{\mathsf{\hat{p}}^{-2j-1} \mathsf{\gamma_{\hat{p}}^{k+1}}}{q'} \varphi(q')
	\label{eq:expand2quantA} 
\end{align}
where, $P_n^{(Q)}(q|q')$ is given by Eq. \eqref{eq:basisFactors} and the superscript $(Q)$ refers the to quantization rule used. The momentum kernel $\mel{q}{\mathsf{\hat{p}}^{-2j-1} \mathsf{\gamma_{\hat{p}}^{k+1}}}{q'} $ is evaluated by inserting the resolution of the identity $\mathsf{1}=\int_{-\infty}^{\infty} dp \ket{p}\bra{p}$, and using the plane wave expansion $\braket{q}{p}=e^{iqp/\hbar}/\sqrt{2\pi\hbar}$, i.e.,
\begin{align}
	\mel{q}{\mathsf{\hat{p}}^{-2j-1} \mathsf{\gamma_{\hat{p}}^{k+1}}}{q'} 
	=& \int_{-\infty}^{\infty} \frac{dp}{2\pi\hbar} \exp[\frac{i}{\hbar}(q-q')p] \frac{1}{p^{2j+1}} \left(  \sqrt{1 + \frac{p^2}{\mu_o^2 c^2}} \right)^{k+1}.
	\label{eq:pkernelGen}
\end{align}
The integral on the right hand side of Eq. \eqref{eq:pkernelGen} diverges because of the pole with order $2j+1$ at $p=0$. Moreover, it has branch points at $\pm i\mu_o c$ for even values of $k$. Now, this has already been evaluated \cite{flores2022relativistic} for the case when $j=k=0$ and can be similarly evaluated as a distributional Fourier transform using the contours shown in Fig. \ref{fig:contour_pos}. The evaluation of Eq. \eqref{eq:pkernelGen} is done by taking its complex extension and taking the average of the integrals $\int_{\Gamma^\pm} dz f(z)z^{-2j-1}$, where the paths $\gamma^+$ ($\gamma^-$) passes above (below) the pole at $z=0$. Performing this integration assigns a value to Eq. \eqref{eq:pkernelGen} which coincides with the Hadamard finite part \cite{Galapon2016}, and is explicitly given as
\begin{align}
	\mel{q}{\mathsf{\hat{p}}^{-2j-1} \mathsf{\gamma_{\hat{p}}^{k+1}}}{q'}  = -\dfrac{1}{2i\hbar} ( f_{j,k}(q,q') + g_{j,k}(q,q')) \text{sgn}(q-q') 
	\label{eq:pkernelSUPP}
\end{align}
where,
\begin{align}
	f_{j,k}(q,q') =&  \dfrac{1}{(2j)!} \left(\dfrac{i}{\hbar} (q-q') \right)^{2j} \int_0^\infty dy e^{-y} \oint_R \dfrac{dz}{2\pi i} \dfrac{1}{z} \sqrt{1 + \dfrac{z^2}{\mu_o^2c^2}}^{k+1} \left( 1 - i\dfrac{\hbar}{q-q'} \dfrac{y}{z}\right)^{2j}
	\label{eq:pkernelevalSUPP} \\
	g_{j,k}(q,q') =& \dfrac{(-1)^j i^k}{(\mu_o c)^{2j}} \left(\dfrac{1-(-1)^{k+1}}{2}\right) \dfrac{2}{\pi} \int_1^\infty dy \exp[- \dfrac{\mu_o c}{\hbar} \abs{q-q'} y] \dfrac{\sqrt{y^2-1}^{k+1}}{y^{2j+1}}.
	\label{eq:pkerneleval1SUPP}
\end{align}
The function $f_{j,k}(q,q')$ is the contribution of the resiude $z=0$ and is rewritten in integral form using the residue theorem, wherein, the contour $R$ is a circle in the complex plane with radius $r<\mu_o c$. Meanwhile, $g_{j,k}(q,q')$ is the contribution of the branch cut which vanishes for odd values of $k$. Thus, the relativistic TOA-operator Eq. \eqref{eq:expand2quantA} now has the expansion
\begin{align}
	(\mathsf{\hat{T}_c}\varphi)(q)= \int_{-\infty}^{\infty} dq'  \dfrac{\mu_o}{i\hbar}  T^{\{Q\}}(q,q') \text{sgn}(q-q') \varphi(q')
\end{align}
where, $T^{\{Q\}}(q,q')$ is the relativistic TKF and has the expansion
\begin{align}
	T^{\{Q\}}(q,q')=\dfrac{1}{2} \sum_{j=0}^\infty & \sum_{k=0}^j \binom{-\frac{1}{2}}{j} \binom{j}{k} \dfrac{(2\mu_o)^j}{(2\mu_o c^2)^{j-k}} ( f_{j,k}(q,q') + g_{j,k}(q,q')) \sum_{n=1}^\infty a_n^{(2j-k)}P_{2j-k}^{\{Q\}}(q|q')  .
	\label{eq:expand2quantc}
\end{align}
An integral form factor for Eq. \eqref{eq:expand2quantc} is obtained by series rearrangement and using the identities in Eq. \eqref{eq:formfactor}.

\subsection*{Modified Weyl-ordered TOA operator}
\label{subsec:weyl}

\noindent Performing the summation yields
\begin{align}
	T^{\{W\}}(q,q') = \dfrac{1}{2} \int_0^{\frac{q+q'}{2}}  ds \mathsf{W}_s(q,q')
	\label{eq:timekernelfunction}
\end{align}
where, 
\begin{align}
	\mathsf{W}_s(q,q') =\mathsf{W}_s^{(1)}(q,q') + \dfrac{2}{\pi}\int_1^\infty dz \exp[-\dfrac{\mu_o c}{\hbar}\abs{q-q'}z] \dfrac{\sqrt{z^2-1}}{z} \mathsf{W}_{s,z}^{(2)}(q,q') 
	\label{eq:timekernelfunction1}
\end{align} 
in which
\begin{align}
	\mathsf{W}_s^{(1)}(q,q') =  \int_0^\infty dy e^{-y} &\oint_R \dfrac{dz}{2\pi i}  \dfrac{1}{z} \sqrt{1 + \dfrac{z^2}{\mu_o^2 c^2}} \nonumber \\
	&\times {_0}F_1 \left[ ; 1; \dfrac{\mu_o V_s^{(W)}(q,q')}{2\hbar^2}  \left( \left(q-q'\right) - i\hbar \dfrac{y}{z} \right)^2 \mathsf{P_W}(s,z,q,q')\right]
	\label{eq:tkf1}
\end{align}
\begin{align}
	\mathsf{W}_{s,z}^{(2)}(q,q') =& \dfrac{1}{2}\left[ 1 - \dfrac{1}{z^2} \left(\dfrac{V_s^{(W)}(q,q')}{\mu_o c^2}\right)^2 + 2i \dfrac{\sqrt{z^2-1}}{z^2} \left( \dfrac{V_s^{(W)}(q,q')}{\mu_o c^2} \right)\right]^{-1/2} + f_{i\rightarrow-i}
	\label{eq:tkf2}
\end{align}
\begin{align}
	\mathsf{P_W}(s,z,q,q') =& \left( \sqrt{1 + \dfrac{z^2}{\mu_o^2 c^2}} + \dfrac{V_s^{(W)}(q,q')}{2\mu_o c^2} \right)
	\label{eq:factor}
\end{align}
\begin{equation}
	V_s^{(W)}(q,q') = V\left( \frac{q+q'}{2} \right) - V(s).
	\label{eq:Vweyl}
\end{equation}
The factor ${_0}F_1(;a;z)$ in Eq. \eqref{eq:tkf1} is a specific hypergeometric function, and the contour $R$ is a circle of radius $r<\mu_o c$ that encloses the pole at $z=0$, while $f_{i\rightarrow-i}$ denotes changing $i$ to $-i$ of the first term in Eq. \eqref{eq:tkf2}. The TKF given by Eqs. \eqref{eq:timekernelfunction}-\eqref{eq:Vweyl} reduces to the known kernel for Weyl-quantized non-relativistic TOA operator Eq. \eqref{eq:nonrelWeyl} in the limit $c\rightarrow\infty$. See Appendix \ref{sec:TKFlimit} for details. 

\subsection*{Modified Born-Jordan-ordered TOA-operator}
\label{subsec:BJ}

\noindent Repeating the same steps yields
\begin{align}
	T^{\{BJ\}}(q,q') =& \dfrac{1}{2(q-q')} \int_0^{q}  ds \int_0^s du \mathsf{B}_{s,u}(q,q') - \dfrac{1}{2(q-q')} \int_0^{q'}  ds \int_0^s du \mathsf{B}_{s,u}(q,q')
	\label{eq:timekernelfunctionBJ}
\end{align}
where, 
\begin{align}
	\mathsf{B}_{s,u}(q,q') =\mathsf{B}_{s,u}^{(1)}(q,q') + \dfrac{2}{\pi}\int_1^\infty dz \exp[-\dfrac{\mu_o c}{\hbar}\abs{q-q'}z] \dfrac{\sqrt{z^2-1}}{z} \mathsf{B}_{s,u,z}^{(2)}(q,q') 
	\label{eq:timekernelfunctionBJ1}
\end{align} 
in which
\begin{align}
	\mathsf{B}_{s,u}^{(1)}(q,q') =\int_0^\infty dy e^{-y} &\oint_R \dfrac{dz}{2\pi i}  \dfrac{1}{z} \sqrt{1 + \dfrac{z^2}{\mu_o^2 c^2}} \nonumber \\
	&\times {_0}F_1 \left[ ; 1; \dfrac{\mu_o V_s^{(BJ)}(u)}{2\hbar^2}  \left( \left(q-q'\right) - i\hbar \dfrac{y}{z} \right)^2 \mathsf{P_{BJ}}(s,u,z,q,q')\right]
	\label{eq:tkfBJ1}
\end{align}
\begin{align}
	\mathsf{B}_{s,u,z}^{(2)}(q,q') =& \dfrac{1}{2}\left[ 1 - \dfrac{1}{z^2} \left(\dfrac{V_s^{(BJ)}(u)}{\mu_o c^2}\right)^2 + 2i \dfrac{\sqrt{z^2-1}}{z^2} \left( \dfrac{V_s^{(BJ)}(u)}{\mu_o c^2} \right)\right]^{-1/2}  + f_{i\rightarrow-i}
	\label{eq:tkfBJ2} 
\end{align}
\begin{align}
	\mathsf{P_{BJ}}(s,u,z,q,q') =& \left( \sqrt{1 + \dfrac{z^2}{\mu_o^2 c^2}} + \dfrac{V^{(BJ)}(s,u)}{2\mu_o c^2} \right)
	\label{eq:factorBJ}
\end{align}
\begin{equation}
	V^{(BJ)}(s,u) = V(s) - V(u). 
	\label{eq:Vbj}
\end{equation}
The TKF given by Eqs. \eqref{eq:timekernelfunctionBJ}-\eqref{eq:Vbj} also reduces to the known kernel for Born-Jordan quantized non-relativistic TOA operator Eq. \eqref{eq:nonrelBJ} in the limit $c\rightarrow\infty$. 

\subsection*{Modified simple-symmetric-ordered TOA-operator}
\label{subsec:SS}

\noindent Last, we have 
\begin{align}
	T^{\{SS\}}(q,q') =& \dfrac{1}{4} \int_0^{q}  ds \mathsf{S}(s,q) + \dfrac{1}{4} \int_0^{q'}  ds \mathsf{S}(s,q') 
	\label{eq:timekernelfunctionSS}
\end{align}
where
\begin{align}
	\mathsf{S}(s,x) =\mathsf{S}^{(1)}(s,x) + \dfrac{2}{\pi}\int_1^\infty dz \exp[-\dfrac{\mu_o c}{\hbar}\abs{q-q'}z] \dfrac{\sqrt{z^2-1}}{z} \mathsf{S}_{z}^{(2)}(s,x) 
	\label{eq:timekernelfunctionSS1}
\end{align} 
in which
\begin{align}
	\mathsf{S}^{(1)}(s,x) =\int_0^\infty dy e^{-y} & \oint_R \dfrac{dz}{2\pi i}  \dfrac{1}{z} \sqrt{1 + \dfrac{z^2}{\mu_o^2 c^2}} \nonumber \\
	&\times {_0}F_1 \left[ ; 1; \dfrac{\mu_o V^{(SS)}(s,x)}{2\hbar^2}  \left( \left(q-q'\right) - i\hbar \dfrac{y}{z} \right)^2 \mathsf{P_{SS}}(s,z,x)\right]
	\label{eq:tkfSS1}
\end{align}
\begin{align}
	\mathsf{S}_{z}^{(2)}(s,x) =& \dfrac{1}{2}\left[ 1 - \dfrac{1}{z^2} \left(\dfrac{V^{(SS)}(s,x)}{\mu_o c^2}\right)^2 + 2i \dfrac{\sqrt{z^2-1}}{z^2} \left( \dfrac{V^{(SS)}(s,x)}{\mu_o c^2} \right)\right]^{-1/2} + f_{i\rightarrow-i}
	\label{eq:tkfSS2}
\end{align}
\begin{align}
	\mathsf{P_{SS}}(s,z,x) =& \left( \sqrt{1 + \dfrac{z^2}{\mu_o^2 c^2}} + \dfrac{V^{(SS)}(s,x)}{2\mu_o c^2} \right)
	\label{eq:factorSS}
\end{align}
\begin{equation}
	V^{(SS)}(s,x) = V(x) - V(s).
	\label{eq:Vss}
\end{equation}
The TKF given by Eqs. \eqref{eq:timekernelfunctionSS}-\eqref{eq:Vss} also reduces to the known kernel for simple-symmetric quantized non-relativistic TOA-operator Eq. \eqref{eq:nonrelSS} in the limit $c\rightarrow\infty$. 

In general, a closed form expression for the relativistic TKFs $T^{\{W\}}(q,q')$, $T^{\{BJ\}}(q,q')$, and $T^{\{SS\}}(q,q')$ may be intractable because of how we assigned a finite value to the divergent integral Eq. \eqref{eq:pkernelSUPP}. It is possible that a tractable form may be obtained using a different assignment to the divergent integral. However, we justify the use of $T^{\{W\}}(q,q')$, $T^{\{BJ\}}(q,q')$, and $T^{\{SS\}}(q,q')$ because it reduces to the non-relativistic time kernel\cite{galapon2018quantizations}.

\section{Barrier traversal time operator}
\label{sec:timeopr}

\begin{figure}[t!]
	\centering
	\includegraphics[width=0.75\textwidth]{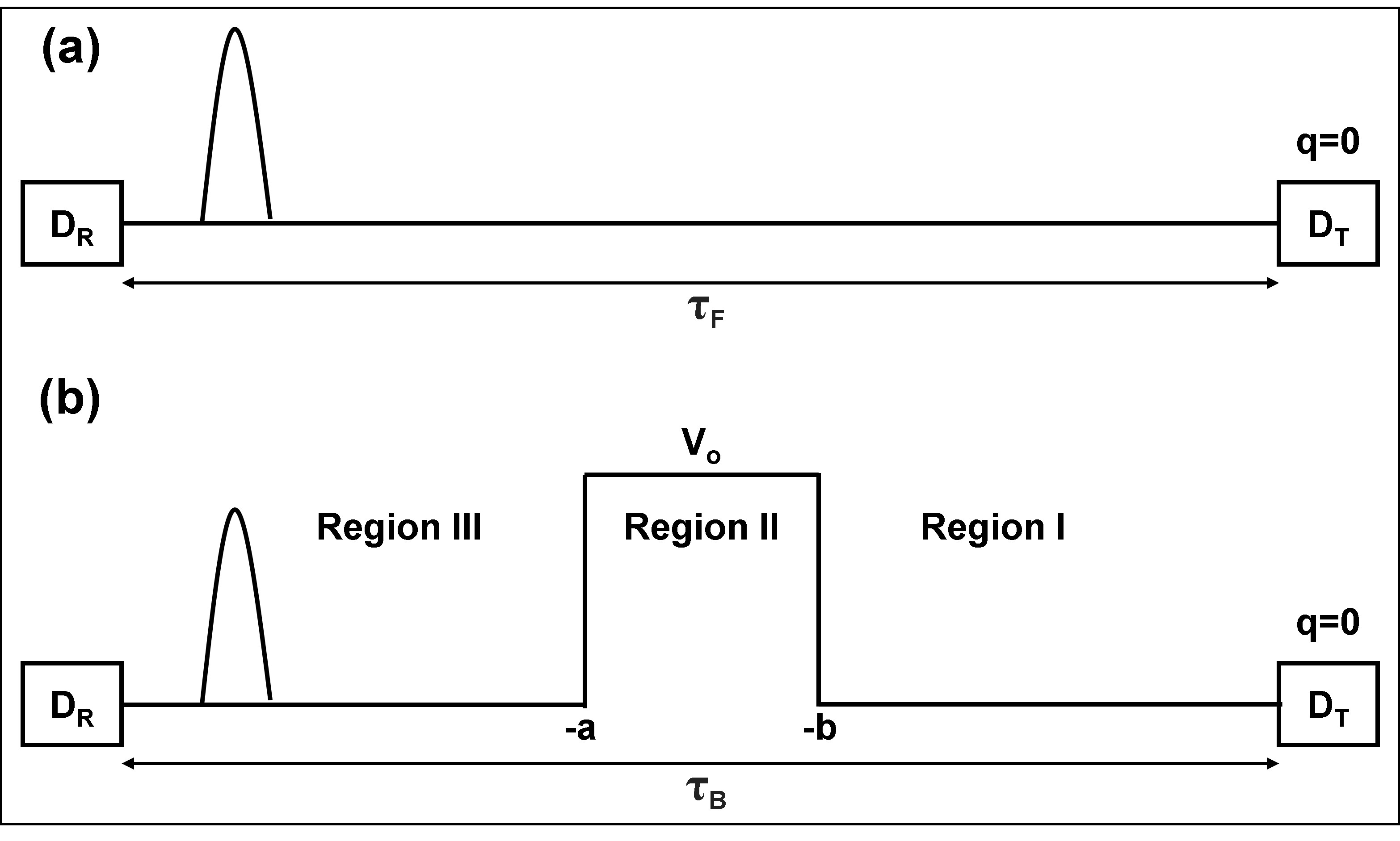}
	\caption{Measurement scheme for the traversal time of a particle in the (a) absence of a barrier, and (b) presence of a barrier. The wavepacket $\psi_o(q)$ is prepared between the detectors $D_R$ and $D_T$ such that its tails does not extend to the barrier region. }
	\label{fig:measurement} 
\end{figure}

We use the measurement scheme shown in Fig. \ref{fig:measurement}. Two detectors $D_T$ and $D_R$ are placed at the arrival point $q=0$ and in the far left, respectively. A square potential barrier of height $V(q)=V_o>0$ and length $L=a-b$ is then placed between the detectors $a<q<b<0$. Next, a wavepacket $\psi(q)$ initially centered at $q=q_o$ with momentum $p_o$ is placed between $D_R$ and the barrier such that the tail of $\psi(q)$ does not initially `leak' into the barrier. The wavepacket is then launched at $t=0$ towards $D_T$ which records the arrival of the particle while the detector $D_R$ does not record any data. This is done to avoid altering the propagation of $\psi(q)$ and provide an indirect but accurately realistic way of obtaining the TOA of the particle at the origin \cite{PhysRevLett.108.170402,sombillo2014quantum,sombillo2018barrier}. The same measurement scheme is employed in the absence of the barrier. 

The measurement is repeated several times for an ensemble of identically prepared particles to obtain a TOA distribution at $D_T$. We assume that the measured TOA distribution has an ideal distribution generated by the spectral resolution of a corresponding TOA-operator $\mathsf{\hat{T}_F}$ and $\mathsf{\hat{T}_B}$ in the absence and presence of the potential barrier, respectively. In the succeeding expressions, the subscript $F$ ($B$) will indicate the case when the barrier is absent (present). The traversal time across the barrier is then deduced from the difference of the average value of the measured TOA 
\begin{equation}
	\Delta\bar{\tau}=\bar{\tau}_F-\bar{\tau}_B = \mel{\psi}{\mathsf{\hat{T}_F}}{\psi} - \mel{\psi}{\mathsf{\hat{T}_B}}{\psi}
\end{equation}
and is assumed to be the expectation value of the TOA-operator.

In the absence of the barrier, the relativistic TKFs $T^{\{W\}}_F(q,q')$, $T^{\{BJ\}}_F(q,q')$, and $T^{\{SS\}}_F(q,q')$ is obtained by substituting $V(q)=0$ into Eqs. \eqref{eq:timekernelfunction}, \eqref{eq:timekernelfunctionBJ} and \eqref{eq:timekernelfunctionSS}, respectively. All ordering rules will yield the same TKF
\begin{equation}
	\tilde{T}_F(\eta,\zeta) = \frac{\eta}{2} \mathsf{T}_F(\zeta),
	\label{eq:freeFactor}
\end{equation}
where, 
\begin{equation}
	\mathsf{T}_F(\zeta) = 1 + \dfrac{2}{\pi} \int_1^\infty dz \dfrac{\sqrt{z^2-1}}{z} \exp(-\dfrac{\mu_o c}{\hbar}\abs{\zeta}z).
	\label{eq:freefactor}
\end{equation}
The operator corresponding to the TKF $\tilde{T}_F(\eta,\zeta)$ coincides with the Rigged Hilbert space extension of Razavi's relativistic free TOA-operator \cite{flores2022relativistic} 
\begin{align}
	(\mathsf{\hat{T}_\text{Ra}} \phi)(q) = \int_{-\infty}^\infty dq'\mel{q}{\mathsf{\hat{T}_\text{Ra}}}{q'} \phi(q')
	\label{eq:RHS_Ra}
\end{align}
wherein, it was shown that the physical quantities associated with Eq. \eqref{eq:RHS_Ra} are consistent with special relativity. Now, the TKFs $T^{\{W\}}(q,q')$, $T^{\{BJ\}}(q,q')$, and $T^{\{SS\}}(q,q')$ were derived under the assumption that the interaction potential $V(q)$ is analytic. However, it can still be applied to piecewise potentials such as the square barrier because the TKFs are in integral form. We will justify this assumption later by establishing that in the classical limit $\hbar\rightarrow0$, the operator for the square potential barrier corresponding to the TKFs reduce to its ``classical'' relativistic TOA. 

The TKF using the modified Wey ordering Eq. \eqref{eq:timekernelfunction} may be obtained by mapping the potential $V(q)$ from $(q,q')$ coordinates into three non-overlapping regions in the $(\eta,\zeta)$ coordinate wherein $\eta=(q+q')/2$ and $\zeta=q-q'$. In this coordinate system, the arrival point is now at $\eta=0$ and $V(\eta)=V_o$ for $a<\eta<b<0$ and zero outside the interval $(a,b)$. For Region I, it is easy to see that $V(\eta)=0$ for the entire integration region of Eq. \eqref{eq:timekernelfunction}. Meanwhile, for Region II, we now have $V(\eta)=V_o$ and it is necessary to split the integral Eq. \eqref{eq:timekernelfunction} into two parts as $V(s)=0$ for $b<s<0$ while $V(s)=V_o$ for $\eta<s<b$. Last, for Region III we have $V(\eta)=0$ and split the integral Eq. \eqref{eq:timekernelfunction} into three parts as $V(s)=V_o$ for $a<s<b$ while $V(s)=0$ outside this interval. Performing these operations will yield
\begin{align}
	\tilde{T}_B^{(I)}(\eta,\zeta)=&  \dfrac{\eta}{2} \mathsf{T}_F(\zeta) \nonumber \\
	\tilde{T}_B^{(II)}(\eta,\zeta)=& \left(\dfrac{\eta+b}{2}\right)\mathsf{T}_F(\zeta)-\dfrac{b}{2}\mathsf{T}_B(V_o,\zeta) \label{eq:barrierkernel} \\
	\tilde{T}_B^{(III)}(\eta,\zeta) =& \left(\dfrac{\eta+L}{2}\right)\mathsf{T}_F(\zeta)-\dfrac{L}{2}\mathsf{T}_B(-V_o,\zeta) \nonumber 
\end{align}
in which $\mathsf{T}_F(\zeta)$ is given by Eq. \eqref{eq:freefactor} and
\begin{align}
	\mathsf{T}_B(V_o,\zeta) =& \mathsf{F}_B(V_o,\zeta) + \dfrac{2}{\pi}\int_1^\infty dz \exp[-\dfrac{\mu_o c}{\hbar} \abs{\zeta}z] \dfrac{\sqrt{z^2-1}}{z}\mathsf{G}_B(V_o,z). \label{eq:barrierfactor} \\
	\mathsf{F}_B(V_o,\zeta) =& \int_0^\infty dy e^{-y} \oint_R \dfrac{dz}{2\pi i}  \dfrac{1}{z} \sqrt{1 + \dfrac{z^2}{\mu_o^2 c^2}} {_0}F_1 \left[ ; 1; \dfrac{\mu_o V_o}{2\hbar^2}  \left( \zeta - i\hbar \dfrac{y}{z} \right)^2 \left( \sqrt{1+\dfrac{z^2}{\mu_o^2c^2}} + \dfrac{V_o}{2\mu_o c^2} \right)\right]
	\label{eq:bkfF} \\
	\mathsf{G}_B(V_o,z) =& \dfrac{1}{2}\left\{ \left[ 1 - \dfrac{1}{z^2} \left(\dfrac{V_o}{\mu_o c^2}\right)^2 + 2i \dfrac{\sqrt{z^2-1}}{z^2} \left( \dfrac{V_o}{\mu_o c^2} \right)\right]^{-1/2} + g_{i\rightarrow-i} \right\}. \label{eq:bkfG}
\end{align}

We now work in the original $(q,q')$-coordinate of our system to evaluate the TKF $T^{\{BJ\}}(q,q')$ given by Eqs. \eqref{eq:timekernelfunctionBJ}-\eqref{eq:factorBJ} and later transform to the coordinates $(\eta,\zeta)$. For Region I, $V(q)=0$ for the entire integration region of Eq. \eqref{eq:timekernelfunctionBJ}. Meanwhile, for Region II, it is necessary to split the integral over $u$ of Eq. \eqref{eq:timekernelfunctionBJ} into two parts as $V(u)=0$ for $b<u<0$ while $V(u)=V_o$ for $s<u<b$ while $V(s)=V_o$ over the whole region of $s$. We again repeat the same steps for Region III, and split the integral over $u$ of Eq. \eqref{eq:timekernelfunctionBJ} into three parts as $V(u)=V_o$ for $a<u<b$ while $V(u)=0$ outside this interval. Then, $V(s)=V_o$ over the whole region of $s$ in Eq. \eqref{eq:timekernelfunctionBJ}. Performing these operations and transforming into the coordinates $(\eta,\zeta)$ will yield the same TKFs as Eq. \eqref{eq:barrierkernel}. Repeating the same  procedure will  yield the the same TKFs $T^{\{SS\}}(q,q')$. In the succeeding discussion, we shall only refer to the modified Weyl-ordered operator since the same results will also hold for the Born-Jordan and simple-symmetric case.

\section{Classical limit of the free and barrier TKFs}
\label{sec:ClassLim}

We now prove that the TKFs corresponding to the TOA-operator for the free and barrier case are indeed the quantization of the CRTOA by taking their inverse Weyl-Wigner transform 
\begin{equation}
	\tilde{t}(q_o,p_o) = \dfrac{\mu_o}{i\hbar}\int_{-\infty}^{\infty} d\zeta e^{-ip_o\zeta/\hbar} \tilde{T}(q_o,\zeta)\text{sgn}(\zeta)
	\label{eq:Weyl-Wigner}
\end{equation}
where, $q_o$ and $p_o$ are the initial position and momentum, respectively. For the free case, this is done by substituting Eq. \eqref{eq:freeFactor} to Eq. \eqref{eq:Weyl-Wigner} which yields
\begin{align}
	\tilde{t}_F = \dfrac{\mu_o}{i\hbar} \dfrac{q_o}{2} \int_{-\infty}^{\infty} d\zeta e^{-ip_o\zeta/\hbar} \text{sgn}(\zeta)
	+ \dfrac{\mu_o}{i\hbar} \dfrac{q_o}{2}   \dfrac{2}{\pi}\int_1^\infty dz \dfrac{\sqrt{z^2-1}}{z}  \int_{-\infty}^{\infty} d\zeta \exp[-\dfrac{\mu_o c}{\hbar} \abs{\zeta}z]  e^{-ip_o\zeta/\hbar} \text{sgn}(\zeta).
	\label{eq:classkernel}
\end{align}
The first term of Eq. \eqref{eq:classkernel} is evaluated by taking the inverse of the distributional Fourier transform \cite{gel1964ov}
\begin{equation}
	\int_{-\infty}^{\infty} dx x^{-m} e^{i\sigma x} = i^m \dfrac{\pi}{(m-1)!} \sigma^{m-1} \text{sgn}\sigma.
	\label{eq:nonrelb}
\end{equation}
Meanwhile, the order of integration for the second term of Eq. \eqref{eq:classkernel} are interchanged, and the inner integral is evaluated as a Laplace transform. The resulting expression is further evaluated using the integral identity \cite{flores2022relativistic}
\begin{equation}
	\int_1^\infty dz \dfrac{\sqrt{z^2-1}}{z} \dfrac{a^2}{a^2 +b^2z^2} = \dfrac{\pi}{2} \left(-1 + \sqrt{1+\dfrac{a^2}{b^2}}\right). 
\end{equation}
for all real $(a,b)$, which can also be obtained using the
calculus of residues. Thus, the classical limit of the free TOA-operator corresponding to $\tilde{T}_F(\eta,\zeta)$ is
\begin{align}
	\tilde{t}_F =& -\dfrac{\mu_o q_o}{p_o} \sqrt{1 + \dfrac{p_o^2}{\mu_o^2 c^2}},
	\label{eq:freeclass}
\end{align}
which is the known free CRTOA obtained from directly integrating Eq.\eqref{eq:classreltoa0}. 

In the presence of the potential barrier, it easily follows from Eq. \eqref{eq:freeclass} that the classical limit of the TKF for Region I $\tilde{T}_B^{(I)}(\eta,\zeta)$ is $\tilde{t}_B^{(I)}=\tilde{t}_F$. For Region II,  
the Weyl-Wigner transform of the TKF $\tilde{T}_B^{(II)}(\eta,\zeta)$ is  
\begin{align}
	\tilde{t}_B^{(II)} =& - \dfrac{\mu_o (q_o + b)}{p} \sqrt{1 + \dfrac{p_o^2}{\mu_o^2 c^2}} -\dfrac{b}{2} \dfrac{\mu_o}{i\hbar}\int_{-\infty}^{\infty} d\zeta e^{-ip_o\zeta/\hbar} \mathsf{T}_B(V_o,\zeta) \text{sgn}(\zeta), 
	\label{eq:WelWignerRegionII}
\end{align}
wherein
\begin{align}
	\int_{-\infty}^{\infty} & d\zeta e^{-ip_o\zeta/\hbar} \mathsf{T}_B(V_o,\zeta) \text{sgn}(\zeta) \nonumber \\
	=& \int_{-\infty}^{\infty} d\zeta e^{-ip_o\zeta/\hbar} \mathsf{F}_B(V_o,\zeta) \text{sgn}(\zeta)+ \left( \dfrac{2 \hbar}{i p_o} \right) \dfrac{2}{\pi} \int_1^\infty dz \mathsf{G}_B(V_o,z) \dfrac{\sqrt{z^2-1}}{z} \dfrac{p_o^2}{p_o^2 + \mu_o^2 c^2 z^2}.  \label{eq:WelWignerRegionIIa}
\end{align}
The first term of Eq. \eqref{eq:WelWignerRegionIIa} is evaluated by expanding the hypergeometric function in $\mathsf{F}_B(V_o,\zeta)$ using its power series representation to perform a term-by-term integration. The resulting series converges as long the initial energy of the particle is above the barrier height, i.e.
\begin{align}
	\int_{-\infty}^{\infty} & d\zeta e^{-ip_o\zeta/\hbar} \mathsf{F}_B(V_o,\zeta) \text{sgn}(\zeta) \nonumber \\
	=& \dfrac{2\hbar}{i p_o} \sum_{j=0}^{\infty} \dfrac{(2j)!}{j!j!} \left( \dfrac{-\mu_o V_o}{2\hbar^2} \right)^j \sum_{k=0}^j \binom{j}{k} \left( \dfrac{V_o}{2\mu_o c^2} \right)^{j-k} \nonumber \\
	&\times\left\{ \sqrt{1 + \dfrac{p_o^2}{\mu_o^2 c^2}}^{k+1} -\left(\dfrac{p_o^2}{\mu_o^2c^2}\right)^{j+1} \binom{\frac{k+1}{2}}{j+1} {_2}F_1\left[ 1, \frac{1}{2} + j - \frac{k}{2} ; j+2 ; -\dfrac{p_o^2}{\mu_o^2 c^2} \right]\right\}
	\nonumber \\
	=& \left(\dfrac{2\hbar}{i p_o}\right) \sqrt{1 + \dfrac{p_o^2}{\mu_o^2 c^2}} \left[ 1 + \dfrac{2\mu_o V_o}{p_o^2}\left( \sqrt{1 + \dfrac{p_o^2}{\mu_o^2 c^2}} + \dfrac{V_o}{2\mu_o c^2}  \right) \right]^{-1/2} \nonumber \\
	&- \left(\dfrac{2\hbar}{i p_o}\right) \dfrac{2}{\pi} \int_1^\infty dz \mathsf{G}_B(V_o,z) \dfrac{\sqrt{z^2-1}}{z} \dfrac{p_o^2}{p_o^2 + \mu_o^2 c^2 z^2}
	\label{eq:evalA}
\end{align}
The second line follows from using the integral representation of the Gauss hypergeometric function 
\begin{equation}
	{_2}F_1(\alpha,\beta;\gamma;z) = \dfrac{\Gamma(\gamma)}{\Gamma(\beta)\Gamma(\gamma-\beta)} \int_0^\infty dt \dfrac{t^{c-b-1}(1+t)^{a-c}}{(t+1-z)^a}
	\label{eq:gauss2F1}
\end{equation}
for $Re[\gamma]>\Re[\beta]>0$ and $\abs{\text{Arg}(1-z)}<\pi$. Combining Eqs. \eqref{eq:WelWignerRegionII}-\eqref{eq:evalA} thus yields 
\begin{align}
	\tilde{t}_B^{(II)} 	=& - \dfrac{\mu_o (q_o + b)}{p_o} \sqrt{1 + \dfrac{p_o^2}{\mu_o^2 c^2}}  + \dfrac{b}{c} \sqrt{\dfrac{1+\dfrac{p_o^2}{\mu_o^2c^2}}{\left( \sqrt{1+\dfrac{p_o^2}{\mu_o^2c^2}} + \dfrac{V_o}{\mu_o c^2}\right)^2-1}}. 
\end{align}
The first term of $\tilde{t}_B^{(II)}$ is the free CRTOA from the edge of the barrier to the origin while the second term is the traversal time on top of the barrier. Repeating the same steps, the Weyl-Wigner transform of $\tilde{T}_B^{(III)}(\eta,\zeta)$ is
\begin{align}
	\tilde{t}_B^{(III)} =& - \dfrac{\mu_o (q_o + L)}{p_o} \sqrt{1 + \dfrac{p_o^2}{\mu_o^2 c^2}} + \dfrac{L}{c} \sqrt{\dfrac{1+\dfrac{p_o^2}{\mu_o^2c^2}}{\left( \sqrt{1+\dfrac{p_o^2}{\mu_o^2c^2}} - \dfrac{V_o}{\mu_o c^2}\right)^2-1}}. 
\end{align}
The first term of $\tilde{t}_B^{(III)}$ is the traversal time across the interaction free region while the second term is the traversal time across the barrier region. The Weyl-Wigner transforms $\tilde{t}_B^{(II)}$ and $\tilde{t}_B^{(III)}$ also coincide with CRTOA obtained from directly integrating Eq. \eqref{eq:classreltoa0}. 

In general, the classical limit of the TKF for a given quantization scheme is obtained by 
\begin{equation}
	\tilde{t}(q_o,p_o) = \lim_{\hbar\rightarrow0} \dfrac{\mu_o}{i\hbar}\int_{-\infty}^{\infty} d\zeta e^{-ip_o\zeta/\hbar} \tilde{T}^{\{Q\}}(q_o,\zeta)\text{sgn}(\zeta),
	\label{eq:classlimit}
\end{equation}
wherein the integral is understood in a distributional sense, provided that the limit exists \cite{galapon2018quantizations}. Notice that the Weyl-Wigner transform Eq. \eqref{eq:Weyl-Wigner} does not involve the vanishing of $\hbar$. Now, Eq. \eqref{eq:classlimit} implies that the classical limit of the TKF for a given quantization scheme is, in general, dependent on positive powers of $\hbar$. Such is the case for the Born-Jordan and simple-symmetric ordering. Performing the limit $\hbar\rightarrow0$ then reduces to classical limits of the TKFs $T^{\{BJ\}}(q,q')$ and $T^{\{SS\}}(q,q')$ into that equal to the Weyl-Wigner transform of $T^{\{W\}}(q,q')$. 

\section{Expected Barrier Traversal Time}
\label{sec:expec}

We now assume that the average value of the measured TOA $\bar{\tau}$ at the detector $D_T$ (see Fig. \ref{fig:measurement}) is equal to the expectation value of the operator $\mathsf{\hat{T}}$, i.e.
\begin{align}
	\bar{\tau} =& \mel{\psi}{\mathsf{\hat{T}}}{\psi} = \int_{-\infty}^\infty dq  \psi^*(q)  \int_{-\infty}^\infty dq'   \dfrac{\mu_o}{i\hbar} T(q,q') \text{sgn}(q-q') \psi(q'). \label{eq:expec} 
\end{align}
The incident wavefunction is assumed to be prepared in a pure state $\psi(q)=\varphi(q)e^{i k_o q}$ with momentum expectation value $p_o=\hbar k_o$, where $\mel{\varphi}{\mathsf{\hat{p}}}{\varphi}=0$. We further assume that $\varphi(q)$ is infinitely differentiable and impose the condition that the support of $\varphi(q)$ is in Region III such that the tail of $\varphi(q)$ does not 'leak' into the barrier. To evaluate Eq. \eqref{eq:expec}, it will be convenient to perform a change of variables from $(q,q')$ to $(\eta,\zeta)$ such that $\bar{\tau}=\Im(\bar{\tau}^*)$ wherein $\bar{\tau}^*$ is the complex-valued TOA given by 
\begin{align}
	\bar{\tau}^* = -\dfrac{2\mu_o}{\hbar} \int_{-\infty}^\infty & d\eta  \int_0^\infty d\zeta e^{i k_o \zeta } \tilde{T}(\eta,\zeta) \varphi^*\left(\eta - \dfrac{\zeta}{2}\right) \varphi\left(\eta + \dfrac{\zeta}{2}\right). \label{eq:complexTOA}
\end{align}
In the succeeding expressions, we indicate complex-valued quantities with an asterisk $^*$ wherein the imaginary component corresponds to the physical quantity.  

In the absence of the barrier, it easily follows from Eqs. \eqref{eq:freeFactor} and \eqref{eq:complexTOA} that the complex-valued free TOA is
\begin{align}
	\bar{\tau}_F^* = -\dfrac{\mu_o}{\hbar}  &\int_0^\infty d\zeta e^{i k_o \zeta } \mathcal{T}_F(\zeta) \int_{-\infty}^\infty d\eta \eta \varphi^*\left(\eta - \dfrac{\zeta}{2}\right) \varphi\left(\eta + \dfrac{\zeta}{2}\right). 
\end{align}
Meanwhile, in the presence of the barrier, we have
\begin{align}
	\bar{\tau}_B^* =& -\dfrac{\mu_o}{\hbar}  \int_0^\infty d\zeta e^{i k_o \zeta } \int_{-\infty}^\infty d\eta \tilde{T}_B^{(III)}(\eta,\zeta)\varphi^*\left(\eta - \dfrac{\zeta}{2}\right) \varphi\left(\eta + \dfrac{\zeta}{2}\right) \nonumber \\
	=&\bar{\tau}_F^*  - \dfrac{\mu_o L}{\hbar}  \int_0^\infty d\zeta e^{i k_o \zeta } (\mathsf{T}_F(\zeta) - \mathsf{T}_B(-V_0,\zeta)) \int_{-\infty}^\infty d\eta \varphi^*\left(\eta - \dfrac{\zeta}{2}\right) \varphi\left(\eta + \dfrac{\zeta}{2}\right) .
\end{align}
The measurable quantity for deducing the barrier traversal time is the TOA difference between the free and barrier case $\Delta\bar{\tau} = \Im(\Delta\bar{\tau}^*) = \Im(\bar{\tau}_F^*-\bar{\tau}_B^*)$,  which is explicitly given as
\begin{align}
	\Delta\bar{\tau}^* =& \dfrac{\mu_o L}{p_o} \left( Q_c^* - R_c^*\right)
	\label{eq:delta-tau}
\end{align}
wherein 
\begin{align}
	Q_c^* =& k_o \int_0^\infty d\zeta e^{i k_o \zeta } \mathsf{T}_F(\zeta) \Phi(\zeta) \label{eq:Qc}\\
	R_c^* =& k_o \int_0^\infty d\zeta e^{i k_o \zeta } \mathsf{T}_B(-V_0,\zeta) \Phi(\zeta) \label{eq:Rc*} \\
	\Phi(\zeta) =& \int_{-\infty}^\infty d\eta \varphi^*\left(\eta - \dfrac{\zeta}{2}\right) \varphi\left(\eta + \dfrac{\zeta}{2}\right).  \label{eq:Phi} 
\end{align}
The complex-valued dimensionless quantities $Q_c^*$ and $R_c^*$ accounts for the contribution of the barrier and relativistic effects on the non-relativistic free TOA $\mu_o L/p_o$. The physical content of the quantities $Q_c$ and $R_c$ are investigated by taking the asymptotic expansion in the high energy limit $k_o\rightarrow\infty$.

It is easy to see that if we substitute Eq. \eqref{eq:freefactor} to Eq. \eqref{eq:Qc}, then it follows that the quantity $(\mu_o L/p_o)Q_c$ is just the expectation value of the free relativistic TOA-operator calculated by Flores and Galapon \cite{PhysRevA.105.062208}. Thus, 
\begin{equation}
	Q_c \sim \sqrt{1 + \dfrac{p_o^2}{\mu_o^2 c^2}}.
	\label{eq:Qc-asymp}
\end{equation}
which is the relativistic correction to the non-relativistic free TOA $\mu_o L/p_o$. Now, the quantity $R_c^*$ is a Fourier integral with respect to the asymptotic parameter $k_o$. We use the same steps outlined in Sec. \ref{sec:timeopr} for the calculation of the Weyl-Wigner transform of the TKF $\tilde{T}_B^{(III)}(\eta,\zeta)$, and perform repeated integration-by-parts to collect powers of $\hbar$. Taking the imaginary part of $R_c^*$ thus yields
\begin{align}
	\Im[R_c^*] \sim& \sum_{m=0}^\infty \Phi^{(2m)}(0) \dfrac{(-1)^m\hbar^{2m}}{p_o^{2m}} \sum_{j=0}^\infty \dfrac{(2j)!}{(1)_j j!}  \left( \dfrac{\mu_o V_o}{2 p_o^2}\right)^j \sum_{k=0}^j \binom{j}{k} \left\{ \left( -\dfrac{V_o}{2\mu_o c^2}\right)^{j-k} \right. \nonumber \\
	&\times  \left.   \sum_{l=0}^j \binom{\frac{k+1}{2}}{l}\binom{2m+2j-2l}{2j-2l} \left( \dfrac{p_o^2}{\mu_o^2 c^2}\right)^l \right\} + \dfrac{2}{\pi} \int_1^\infty dz \dfrac{\left(\frac{p_o^2}{\mu_o^2 c^2}\right)}{z^2 + {\left(\frac{p_o^2}{\mu_o^2 c^2}\right)}} \dfrac{\sqrt{z^2-1}}{z} \mathsf{G}(V_o,z) \\
	\sim& \sum_{j=0}^\infty \dfrac{(2j)!}{(1)_j j!} \left( \dfrac{\mu_o V_o}{2 p_o^2}\right)^j \sum_{k=0}^j \binom{j}{k} \left( -\dfrac{V_o}{2\mu_o c^2}\right)^{j-k} \left\{  \sqrt{1 + \dfrac{p_o^2}{\mu_o^2 c^2}}^{k+1} \right. \nonumber \\
	&- \left.  \left(\dfrac{p_o^2}{\mu_o^2c^2}\right)^{j+1} \binom{\frac{k+1}{2}}{j+1} {_2}F_1\left[ 1, \frac{1}{2} + j - \frac{k}{2} ; j+2 ; -\dfrac{p_o^2}{\mu_o^2 c^2} \right] \right\}\nonumber \\
	&+ \dfrac{2}{\pi} \int_1^\infty dz \dfrac{\left(\frac{p_o^2}{\mu_o^2 c^2}\right)}{z^2 + {\left(\frac{p_o^2}{\mu_o^2 c^2}\right)}} \dfrac{\sqrt{z^2-1}}{z} \mathsf{G}(V_o,z)
	\label{eq:physicalmeaningR}
\end{align}
The second line Eq. \eqref{eq:physicalmeaningR} follows from the classical limit $\hbar\rightarrow 0$ in which only the terms with $m=0$ will not vanish, wherein we used the normalization conditions $\Phi(0)=1$. The integral representation of the Gauss hypergeometric function, Eq. \eqref{eq:gauss2F1}, is again used to perform the summation which yields
\begin{equation}
	R_c \sim \dfrac{p_o}{\mu_o c} \sqrt{\dfrac{E_p^2}{(E_p - V_o)^2 - \mu_o^2 c^4}}
	\label{eq:RphysMeaning}
\end{equation}
where $E_p = \sqrt{p^2c^2 + \mu_o^2 c^4}$. Thus, $R_c$ is just the ratio of the energy of the incident particle and its energy above the barrier. This leads us to the interpretation that $R_c$ is the effective index of refraction (IOR) of the barrier with respect to the wavepacket. The same interpretation was made in the non-relativistic case for the square potential barrier and well \cite{PhysRevLett.108.170402,pablico2020quantum}. This implies that the traversal time across the barrier is given by $\bar{\tau}_{\text{trav}}=(\mu_o L/p_o)R_c$. 

\begin{figure}[b!]
	\centering
	\includegraphics[width=0.5\textwidth]{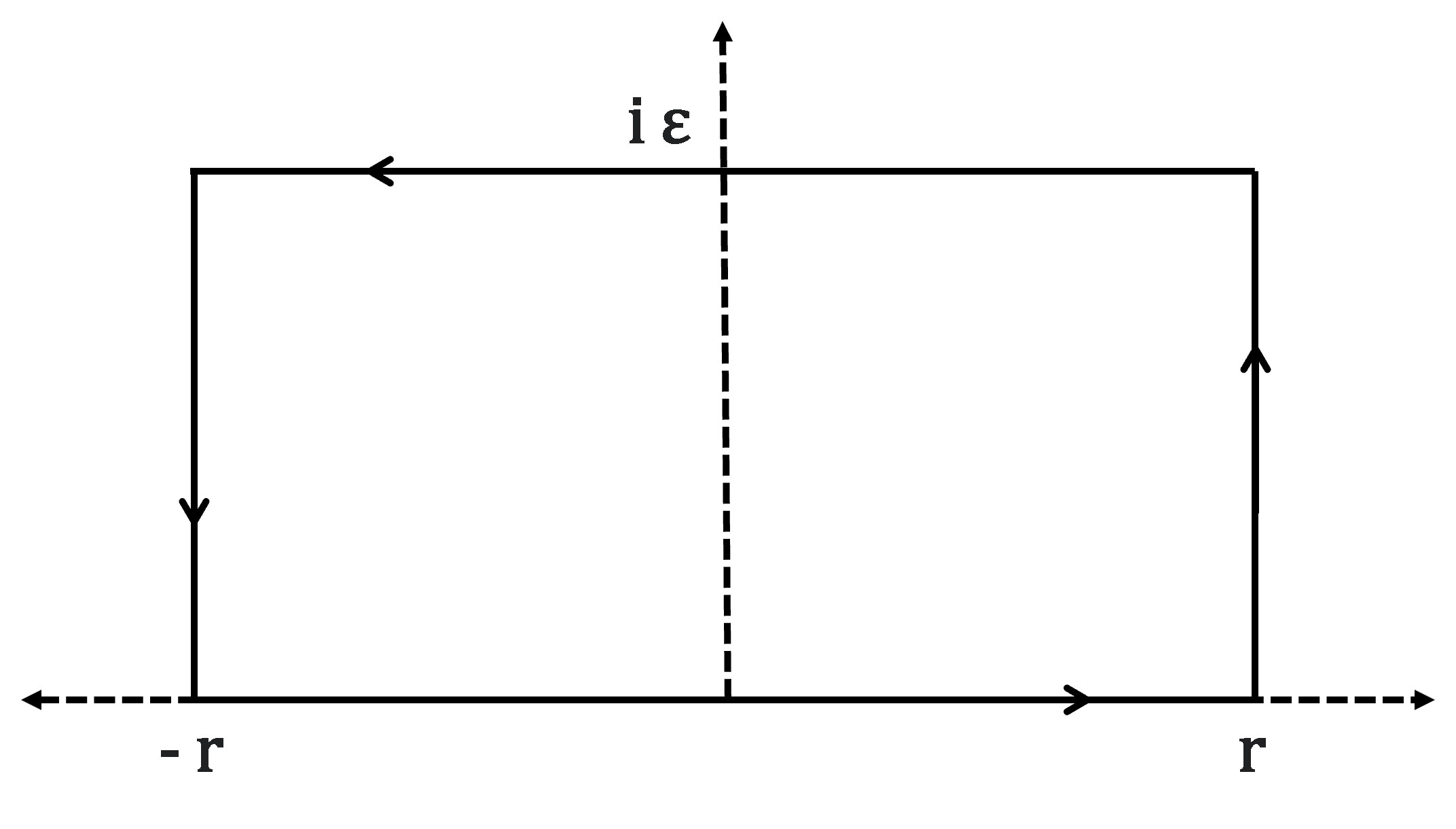}
	\caption{Contour of integration for Eq. \eqref{eq:interchange} leading to the interchange of the order of integration in Eq. \eqref{eq:Rc*fin2}}
	\label{fig:contour1} 
\end{figure}
\begin{figure}[b!]
	\centering
	\includegraphics[width=0.99\textwidth]{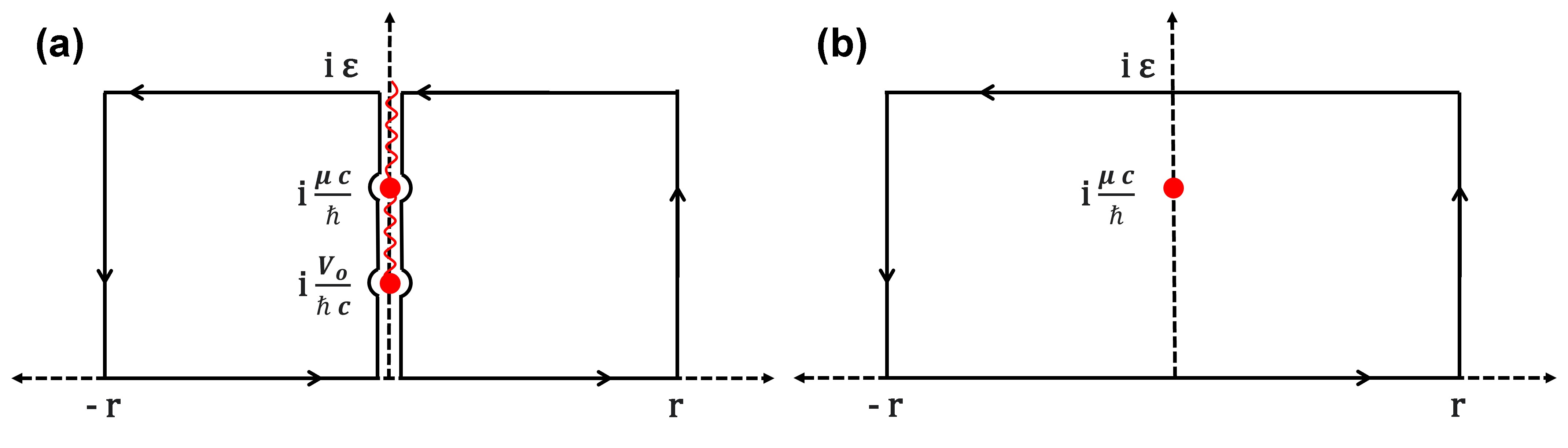}
	\caption{Contours of integration of Eq. \eqref{eq:contour} for the (a) the first integral when $V_o / \hbar c < \mu_o c/\hbar$, and (b) the second integral}
	\label{fig:contourexpecval} 
\end{figure}

We now establish the expected traversal time across the potential barrier and use the same notations as that of Galapon \cite{PhysRevLett.108.170402} for consistency. To evaluate the complex-valued IOR Eq. \eqref{eq:Rc*}, we introduce the inverse Fourier transform of the wavepacket $\varphi(q) = (2\pi)^{-1} \int_{-\infty}^\infty d\tilde{k} e^{i \tilde{k} q} \phi(\tilde{k})$ such that
\begin{equation}
	\Phi(\zeta)=\int_{-\infty}^\infty d\tilde{k} |\phi(\tilde{k})|^2 e^{i \tilde{k} \zeta}.
	\label{eq:PhiV2}
\end{equation}
Substituting Eq. \eqref{eq:PhiV2} to Eq. \eqref{eq:Rc*}, and performing a change of variable $\tilde{k}=k-k_o$ yields
\begin{equation}
	\dfrac{R_c^*}{k_o} = \int_0^\infty d\zeta \mathsf{T}_B(-V_0,\zeta) \int_{-\infty}^\infty dk |\phi(k-k_o)|^2 e^{i k \zeta}
\end{equation}
Notice that $\phi(k-k_o)$ is the Fourier transform of the full incident wavefunction $\psi(q)=e^{i k_o q}\varphi(q)$, i.e.
\begin{equation}
	\phi(k-k_o)=\tilde{\psi}(k)=\dfrac{1}{\sqrt{2\pi}}\int_{-\infty}^\infty dq e^{-ikq}\psi(q)
\end{equation}
Thus, we have
\begin{align}
	\dfrac{R_c^*}{k_o} =\int_0^\infty &d\zeta \mathsf{T}_B(-V_0,\zeta) \int_{-\infty}^\infty dk e^{i k \zeta} \abs{\tilde{\psi}(k)}^2 \label{eq:Rc*fin00} \\
	= \int_0^\infty &d\zeta \mathsf{F}_B(-V_0,\zeta) \int_{-\infty}^\infty dk e^{i k \zeta} \abs{\tilde{\psi}(k)}^2+ \dfrac{2}{\pi} \int_{1}^{\infty} dy  \dfrac{\sqrt{y^2-1}}{y} \mathsf{G}_B(V_o,y) \int_{-\infty}^{\infty} dk  \dfrac{\frac{\mu_o c}{\hbar}y}{k^2 + \frac{\mu_o^2 c^2}{\hbar^2}y^2} \abs{\tilde{\psi}(k)}^2.
	\label{eq:Rc*fin}
\end{align}
The last line follows from interchanging the order of integration in the second term of Eq. \eqref{eq:Rc*fin} but the same cannot be done on the first term. Specifically, if we use the same steps outlined in Sec. \ref{sec:timeopr} to perform a term-by-term integration on the first term of Eq. \eqref{eq:Rc*fin}, then this will lead to an infinite sum of divergent integrals whose values may be assigned using analytic continuation, regularization, and many others. However, it was recently shown by one of us that this naive interchange in the ordering of integrals leading to divergent integrals sometimes miss significant terms \cite{galapon2017problem,tica2019finite}. This was shown to have physical significance in the traversal time of a non-relativistic particle across a potential well \cite{pablico2020quantum}.

\begin{table}[t!]
	\caption{Numerical verification of $\tilde{R}_c$ for spatially narrow Gaussian wavepackets $\sigma = 0.5$ when there are above and below barrier components}
	\centering
	\begin{tabular}{l | c | c |  c}
		\hline
		& Integral: Eq. \eqref{eq:Rc*} & Summation: Eq. \eqref{eq:SumForm} & Evaluated: Eq. \eqref{eq:Rcfinal}\\
		\hline
		\hline 
		$k_o = 2.00 ; V_o = 0.2$ & 1.32442 	& 1.32442 & 1.32442\\
		$k_o = 2.00 ; V_o = 0.3$ & 1.38141 	& 1.38141 & 1.38141\\
		$k_o = 2.00 ; V_o = 0.5$ & --- 		& 1.48255 & 1.48255\\
		$k_o = 2.00 ; V_o = 0.6$ & --- 		& 1.52350 & 1.52350\\
		\hline
		\hline
		$k_o = 0.90 ; V_o = 0.3$ & 0.99882 	& 0.99888 & 0.99888\\
		$k_o = 3.00 ; V_o = 0.3$ & 1.24812 	& 1.24812 & 1.24811\\
		$k_o = 5.00 ; V_o = 0.3$ & 1.09394 	& 1.09394 & 1.09393\\
		\hline
		\hline
		$k_o = 0.15 ; V_o = 0.3$ & 0.18996 	& 0.18996 & 0.18996\\
		$k_o = 0.20 ; V_o = 0.3$ & 0.25253 	& 0.25253 & 0.25253\\
		$k_o = 0.25 ; V_o = 0.3$ & 0.31446	& 0.31446 & 0.31446
		\label{tab:compare}
	\end{tabular}
\end{table}

\begin{table}[t!]
	\caption{Numerical verification of $\tilde{R}_c$ for spatially wide Gaussian wavepackets $\sigma=9.0$ and $V_o=0.3$ when there are only below barrier components }
	\centering
	\begin{tabular}{l | c | c}
		\hline
		& Integral: Eq. \eqref{eq:Rc*} & Evaluated: Eq. \eqref{eq:Rcfinal}\\
		\hline
		\hline 
		$k_o = 0.19$ & $2.23294 \times 10^{-16}$ 	& $1.34410 \times 10^{-29}$\\
		$k_o = 0.25$ & $1.77061 \times 10^{-14}$ 		& $2.01917 \times 10^{-24}$\\
		$k_o = 0.28$ & $1.84479 \times 10^{-16}$ 		& $5.06286 \times 10^{-22}$
		\label{tab:compareBelow}
	\end{tabular}
\end{table}

To make the the interchange in the orders of integration on the the first term of Eq. \eqref{eq:Rc*fin} valid, we use the methods of Pablico and Galapon \cite{pablico2020quantum} and use the contour shown in Fig. \ref{fig:contour1}. We let $p(z)=|\tilde{\psi}(z)|^2$ and assume that $\tilde{\psi}(z)$ does not have any poles in the complex plane, i.e. 
\begin{equation}
	\int_{-\infty}^\infty e^{i x \zeta} p(x) = \int_{-\infty}^\infty dx e^{-(\epsilon-ix)\zeta} p(x+i\epsilon). 
	\label{eq:interchange}
\end{equation} 
This now makes
\begin{align}
	\int_0^\infty &d\zeta \mathsf{F}_B(-V_0,\zeta) \int_{-\infty}^\infty dk e^{i k \zeta} \abs{\tilde{\psi}(k)}^2 =\int_{-\infty}^\infty dk p(k+i\epsilon) \int_{0}^{\infty} d\zeta \mathsf{F}_B(-V_o,\zeta) e^{-(\epsilon - i k)\zeta}.
	\label{eq:Rc*fin2}
\end{align}
The interchange is valid provided that $\epsilon>k$.  We can now use the series representation of the hypergeometric function in $\mathsf{F}_B(-V_o,\zeta)$ and use the same methods outlined in Sec. \ref{sec:timeopr}. This turns the first term of Eq. \eqref{eq:Rc*fin} into
\begin{align}
	\int_0^\infty d\zeta & \mathsf{F}_B(-V_0,\zeta) \int_{-\infty}^\infty dk e^{i k \zeta} \abs{\tilde{\psi}(k)}^2 \nonumber \\
	= i \sum_{n=0}^\infty &\dfrac{(2n)!}{(1)_n n!} \left(\dfrac{\mu_o V_o}{2\hbar^2}\right)^n \int_{-\infty}^\infty dk  p(k+i\epsilon) \text{csgn}(k+i\epsilon)  \nonumber \\
	&\times  \left( \sqrt{1 + \dfrac{\hbar^2 (k+i\epsilon)^2}{\mu_o^2 c^2}} \right)^{n+1} \left( (k+i\epsilon)^2  +  \dfrac{V_o^2}{\hbar^2 c^2}\right)^{-n-\frac{1}{2}} \nonumber \\
	- \dfrac{2i}{\pi} &\int_1^\infty dy  \dfrac{\sqrt{y^2-1}}{y} \mathsf{G}_B(V_o,y) \int_{-\infty}^{\infty} dk p(k+i\epsilon) \dfrac{ \frac{\hbar^2}{\mu^2 c^2} (k+i\epsilon)}{y^2 + \frac{\hbar^2}{\mu^2 c^2}(k+i\epsilon)^2}
	\label{eq:Rc*expandA}
\end{align}
where $\text{csgn}(z)$ is the complex signum function
\begin{align}
	\text{csgn}(z) =
	\begin{cases}
		1 \quad &, \text{Re}(z) > 0 \\
		-1 \quad &, \text{Re}(z) < 0 \\
		\text{sgn}(\text{Im}(z)) \quad &, \text{Re}(z) = 0.
	\end{cases}
\end{align}
To understand the physical content of Eq. \eqref{eq:Rc*expandA}, we consider the following integral in the complex plane, 
\begin{align}
	& \oint dz p(z) \left( \sqrt{1 + \dfrac{\hbar^2 z^2}{\mu_o^2 c^2}} \right)^{n+1} \left( z^2 + \dfrac{V_o^2}{\hbar^2 c^2} \right)^{-n-\frac{1}{2}} \quad \text{and} \quad \oint dz p(z)\dfrac{\frac{\hbar^2}{\mu^2 c^2}z}{y^2 + \frac{\hbar^2}{\mu^2 c^2} z^2},
	\label{eq:contour}
\end{align}
wherein the first integral has four branch points at $z=\{ \pm i \frac{\mu_o c}{\hbar}, \pm i \frac{V_o}{\hbar c}\}$ while the second integral has poles at $z=\pm i\frac{\mu c}{\hbar}$. We assume that the branch points satisfy $V_o/\hbar c < \mu_o c / \hbar$ which is equivalent to the condition $V_o < \mu_o c^2$. The integrals Eq. \eqref{eq:contour} are then evaluated using the contours in Fig. \ref{fig:contourexpecval} (see Appendix \ref{sec:IORdetails} for details) and the resulting expressions are substituted to Eq. \eqref{eq:Rc*fin} which yields 
\begin{align}
	R_c^* =& i \dfrac{\hbar k_o}{\mu c} \int_{0}^\infty  dk  \left( \abs{\tilde{\psi}(k)}^2 - \abs{\tilde{\psi}(- k)}^2\right)  \sqrt{\dfrac{\tilde{E}_k^2}{(\tilde{E}_k-V_o)^2-\mu_o^2 c^4}} \nonumber \\
	&+ k_o \dfrac{2}{\pi} \int_1^\infty dy \dfrac{\sqrt{y^2-1}}{y} \mathsf{G_B}(V_o,y) \int_{-\infty}^\infty dk \abs{\tilde{\psi}(k)}^2 \dfrac{\frac{\mu c}{\hbar} y}{k^2 + \frac{\mu^2 c^2}{\hbar^2} y^2 }
	\label{eq:Rc*b}
\end{align}
in which, $\tilde{E}_k =\sqrt{\hbar^2k^2c^2 + \mu_o^2 c^4}$. It is easy to see that the first term of Eq. \eqref{eq:Rc*b} is generally complex-valued while the second term is always real-valued. Thus, taking the imaginary component of the IOR yields 
\begin{equation}
	\Im[R_c^*] = \dfrac{\hbar k_o}{\mu_o c} \tilde{R}_c = \dfrac{\hbar k_o}{\mu_o c} \text{Re} \left\{ \int_{0}^\infty  dk  \left( \abs{\tilde{\psi}(k)}^2 - \abs{\tilde{\psi}(- k)}^2\right)  \sqrt{\dfrac{\tilde{E}_k^2}{(\tilde{E}_k-V_o)^2-\mu_o^2 c^4}} \right\}
	\label{eq:RcWeak1} 
\end{equation}
The right-hand side of Eq. \eqref{eq:RcWeak1} is only real-valued when $\abs{k}>\kappa_c$, where 
\begin{equation}
	\kappa_c = \sqrt{\dfrac{2\mu_o V_o}{\hbar^2}\left(1+\frac{V_o}{2\mu_o c^2}\right)}
	\label{eq:kappaC}
\end{equation}
provided that $V_o<\mu_o c^2$. Thus,  Eq. \eqref{eq:RcWeak1} becomes
\begin{equation}
	\tilde{R}_c=\tilde{R}_c^{(+)}-\tilde{R}_c^{(-)} = \int_{\kappa_c}^\infty dk  \abs{\tilde{\psi}(+ k)}^2  \sqrt{\dfrac{\tilde{E}_k^2}{(\tilde{E}_k-V_o)^2-\mu_o^2 c^4}} -  \int_{\kappa_c}^\infty dk  \abs{\tilde{\psi}(- k)}^2  \sqrt{\dfrac{\tilde{E}_k^2}{(\tilde{E}_k-V_o)^2-\mu_o^2 c^4}}
	\label{eq:Rcfinal}
\end{equation}

It easily follows that the barrier traversal time now has the form
\begin{equation}
	\bar{\tau}_{\text{trav}} = \dfrac{\mu_o L}{p_o}\Im[R_c^*] = t_c \tilde{R}_c,
	\label{eq:travFinal}
\end{equation}
where, $t_c=L/c$ is the time it takes a photon to traverse the barrier length. The term $\tilde{R}_c^{(+)}$ ($\tilde{R}_c^{(-)}$) characterizes the contribution of the positive (negative) components of the energy distribution of $\tilde{\psi}(k)$ with $\abs{k}>\kappa_c$ to the effective IOR $\tilde{R}_c$. Clearly, the quantity
\begin{equation}
	\bar{\tau}_{\text{trav}}^{(\pm)} = t_c\tilde{R}_c^{(\pm)} = \int_{\kappa_c}^\infty dk \bar{\tau}_{\text{top}}(k) |\tilde{\psi}(\pm k)|^2
\end{equation} 
is the weighted average of the classical above barrier traversal time
\begin{equation}
	\bar{\tau}_{\text{top}}(k) = t_c \sqrt{\dfrac{\tilde{E}_k^2}{(\tilde{E}_k-V_o)^2-\mu_o^2 c^4}} 
\end{equation}
with weights $|\tilde{\psi}(\pm k)|^2$. The effective IOR Eq. \eqref{eq:Rcfinal} shows that the contribution of the below barrier energy components of $\tilde{\psi}(k)$ with $\abs{k}<\kappa_c$ vanishes, which leads us to the same conclusion as that of Galapon \cite{PhysRevLett.108.170402}. That is, the below barrier energy components of $\tilde{\psi}(k)$ are transmitted instantaneously which implies that tunneling, whenever it occurs, is instantaneous. 

Thus, the instantaneous tunneling time predicted in Ref. \cite{PhysRevLett.108.170402} is not a mere consequence of using a non-relativistic theory but is an inherent quantum effect in the context of ``arrival times'' as it still manifests even with a relativistic treatment. However, there is a specific configuration in a tunneling experiment such that this instantaneous tunneling time can be observed. Specifically, it is implied from Eq. \eqref{eq:Rcfinal} that the initial incident wavepacket $\psi(q)$ must be sufficiently spatially wide so that the spread in momentum is narrow. This will ensure that $\tilde{\psi}(k)$ only has below barrier components. Additionally, Eq. \eqref{eq:Rcfinal} rests on the assumption that $\psi(q)$ does not initially `leak' inside the barrier region, as such, the initial incident wavepacket must be placed very far from the barrier.

\begin{figure}[t!]
	\centering
	\includegraphics[width=0.6\textwidth]{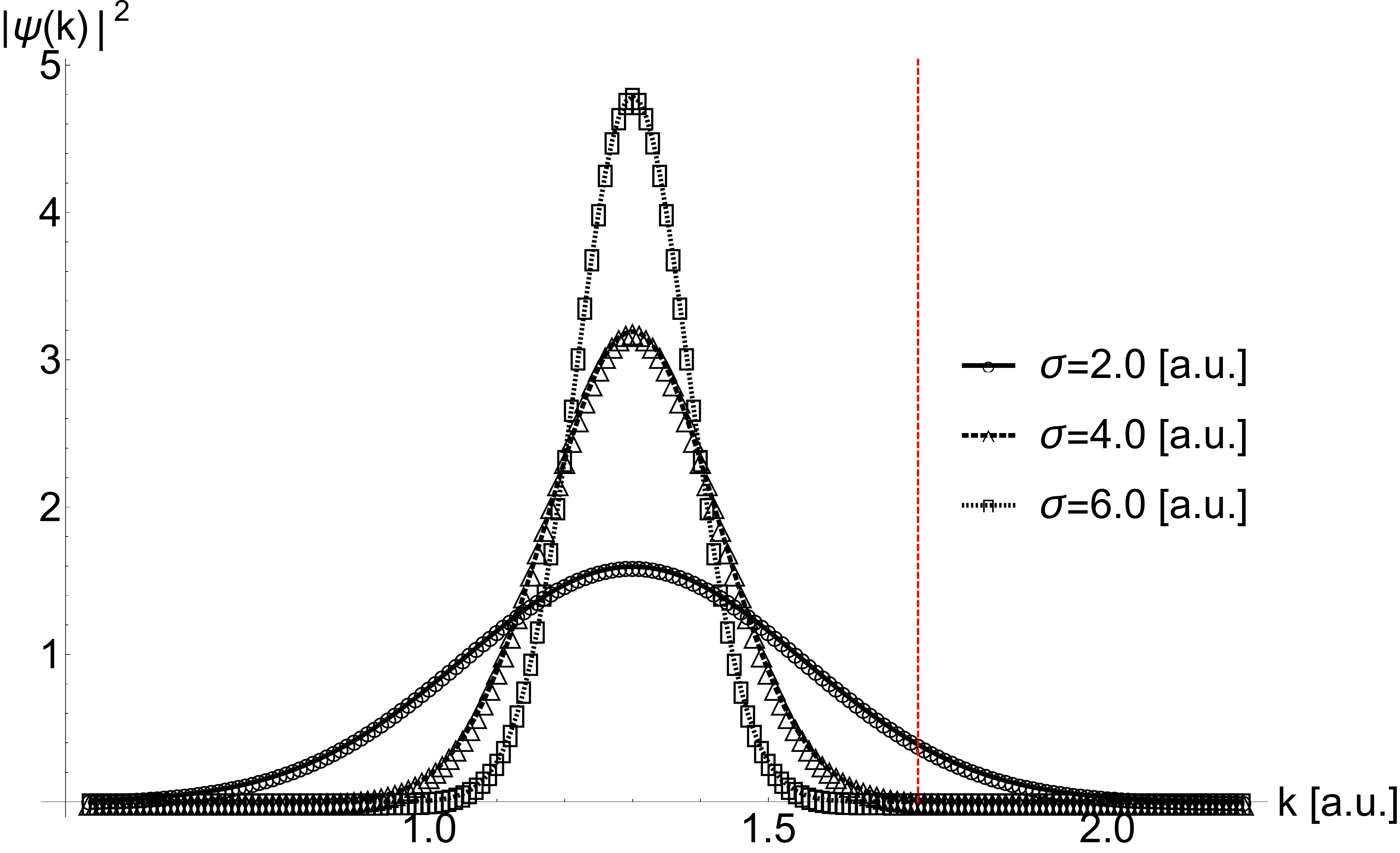}
	\caption{Momentum density distribution $|\tilde{\psi}(k)|^2$ of spatially wide Gaussian wavepackets for the parameters $\mu_o=c=\hbar=1$ with $k_o=1.3$. The red line represents $\kappa_c=1.7025$ with $V_o=0.99$.}
	\label{fig:momentumdist} 
\end{figure}

\section{Barrier traversal time of Gaussian wavepackets}
\label{sec:Gaussian}

We consider an incident Gaussian wavepacket , i.e. 
\begin{equation}
	\varphi(q) = \dfrac{1}{\sqrt{\sigma\sqrt{2\pi}}} \exp[-\dfrac{(q-q_o)^2}{4\sigma^2}]. 
\end{equation}
that is initially centered at $q=q_o$ with a position variance $\sigma^2$. 
In momentum representation, this leads to 
\begin{align}
	\abs{\tilde{\psi}(\pm k)}^2 =& \sqrt{\dfrac{2 \sigma^2}{\pi}} \exp[-2\sigma^2(k\mp k_o)^2]. 
\end{align}
For completeness, we first numerically verify the equivalence of Eqs. \eqref{eq:Rcfinal} and Eq. \eqref{eq:Rc*}. However, Eq. \eqref{eq:Rc*} is numerically taxing and unstable as the potential $V_o$ increases, such that $\mathsf{T}_B(-V_o,\zeta)$ must be represented in an equivalent expression. This is done by using the power series representation of the hypergeometric function in Eq. \eqref{eq:bkfF} to perform a term-by-term integration which yields
\begin{align}
	\mathsf{T}_B(-V_o,\zeta) =&\sum_{l=0}^\infty  \dfrac{(2l)!}{l!l!} \left(\dfrac{-\mu_o V_o}{2\hbar^2}\right)^l \sum_{m=0}^{l}\binom{l}{m}\left(\dfrac{-V_o}{2\mu_o c^2}\right)^{l-m} \sum_{n=0}^l \binom{\frac{m+1}{2}}{n}\left(\dfrac{-\hbar^2}{\mu_o^2 c^2}\right)^n \dfrac{\zeta^{2l-2n}}{(2l-2n)!} \nonumber \\
	&+ \dfrac{2}{\pi}\int_1^\infty dz  \exp[-\dfrac{\mu_o c}{\hbar} \abs{\zeta}z] \dfrac{\sqrt{z^2-1}}{z}\mathsf{G}_B(V_o,z)
	\label{eq:SumForm}
\end{align}
Eq. \eqref{eq:SumForm} is then substituted to Eq. \eqref{eq:Rc*}. This series will converge as long as the initial energy of the particle is above the barrier height for $V_o<\mu_o c^2$. \textcolor{blue}{text}

The equivalent expressions for the effective IOR given by Eqs. \eqref{eq:Rc*}, \eqref{eq:Rcfinal}, and \eqref{eq:SumForm} were numerically evaluated using \textit{Wolfram Mathematica 12 - Student edition}. The computer used has the following specifications: an Intel Core i5-9300H CPU @ 2.40 GHz, 8.0 GB Ram, and a 64-bit operating system $\times$ 64-based processor. Table \ref{tab:compare} compares the values of $\tilde{R}_c$ for spatially narrow Gaussian wavepackets, i.e. the wavepackets have a wide spread in momentum such that it can have both below and above barrier components. The evaluation of Eq. \eqref{eq:Rc*} is numerically taxing for the computer as the potential increases but the equivalent expression Eq. \eqref{eq:SumForm} converges to the same value as that of Eq. \eqref{eq:Rcfinal}. Moreover, it can be seen that for the parameters wherein Eq. \eqref{eq:Rc*} is evaluated, the equivalent expressions Eq. \eqref{eq:SumForm} and \eqref{eq:Rcfinal} all converge to the same value. Table \ref{tab:compareBelow} compares the values of $\tilde{R}_c$ for spatially wide Gaussian wavepackets, i.e. the wavepackets have a narrow spread in momentum such that it only has below barrier components. Eq. \eqref{eq:SumForm} will not converge so we only compare Eqs. \eqref{eq:Rc*} and \eqref{eq:Rcfinal}. It can be seen that the the values become numerically zero, which supports our earlier conclusion. This gives us confidence in the final expression of the barrier traversal time Eq. \eqref{eq:travFinal}.

\begin{figure}[t!]
	\centering
	\includegraphics[width=0.6\textwidth]{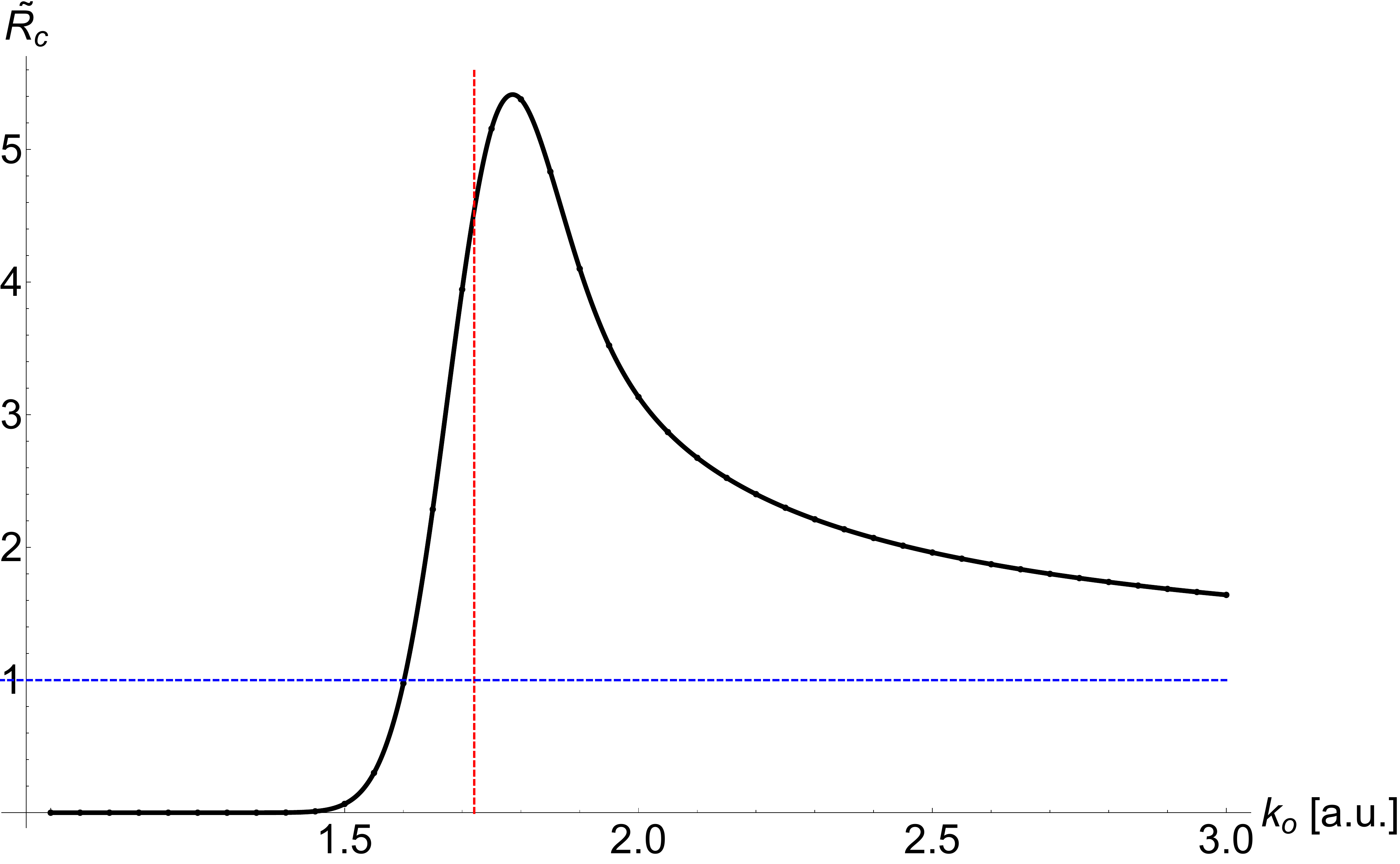}
	\caption{The effective IOR $\tilde{R}_c$ of spatially wide Gaussian wavepackets for the parameters $\mu_o=c=\hbar=1$ with $\sigma=6$. The red line represents $\kappa_c=1.7025$ with $V_o=0.99$. The area below the blue line represents the superluminal region for the traversal time when $\tilde{R}_c < 1$. }
	\label{fig:RcPlot} 
\end{figure}

\begin{figure}[t!]
	\centering
	\includegraphics[width=0.6\textwidth]{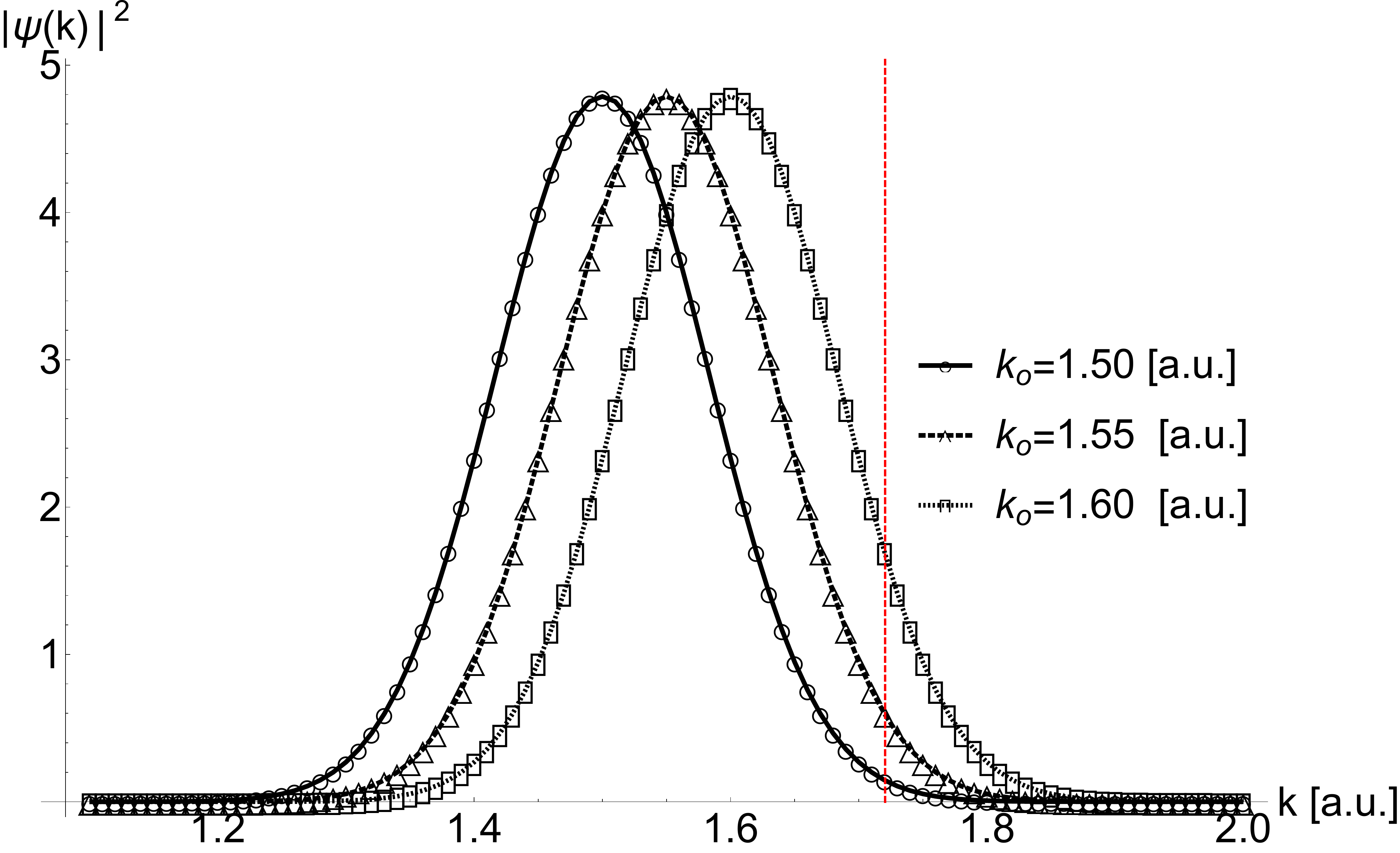}
	\caption{Momentum density distribution $|\tilde{\psi}(k)|^2$ for the parameters $\mu_o=c=\hbar=1$ with $\sigma=6$. The red line represents $\kappa_c=1.7025$ with $V_o=0.99$.}
	\label{fig:momentumdistB} 
\end{figure}

To further appreciate the importance of distinguishing the below and above barrier components, consider Fig. \ref{fig:momentumdist}. The components on the right (left) of the red line $\kappa_c$ are the above (below) barrier components. It can easily be seen from Fig. \ref{fig:momentumdist} that all the components of $|\tilde{\psi}(k)|^2$ for the cases $\sigma=4.0$ and $\sigma=6.0$ are below $\kappa_c$ which will tunnel instantaneously through the barrier $V_o=0.99$. This is easily verified by evaluating Eq. \eqref{eq:Rcfinal} for these parameters, which will yield $\tilde{R}_c\sim0$. 	Fig. \ref{fig:RcPlot} shows the effective IOR $\tilde{R}_c$ for spatially wide Gaussian wavepackets as the initial momentum $k_o$ increases. It can be seen that there is a region where the traversal time $\bar{\tau}_{\text{trav}}$ becomes superluminal as $k_o$ increases such that the spread of $|\tilde{\psi}(k)|^2$ starts to go beyond $\kappa_c$. This is shown in Fig. \ref{fig:momentumdistB}. We can thus estimate that if the initial momentum $k_o<\kappa_c-\sigma_k$, where $\sigma_k$ is the momentum variance, then the traversal time becomes superluminal because $\int_{\kappa_c}^\infty dk |\tilde{\psi}(k)|^2$ is small which effectively leads to $\tilde{R}_c < 1$ or equivalently $\bar{\tau}_{\text{trav}}<t_c$. Moreover, the traversal time becomes subliminal when the initial momentum $k_o>\kappa_c-\sigma_k$, wherein the peak of $\tilde{R}_c$ is roughly at $k_o=\kappa_c+\sigma_k$. The effective IOR $\tilde{R}_c$ then eventually plateaus to some value as all the components of $|\tilde{\psi}(k)|^2$ are above $\kappa_c$.   

\section{Conclusion}
\label{sec:conc}

In this paper, we have given a full account of [\href{https://iopscience.iop.org/article/10.1209/0295-5075/acad9a/meta}{EPL, \textbf{141} (2023) 10001}]. The general form of the quantized relativistic TOA-operators in the presence of an interaction potential were also obtained using a modified Weyl, Born-Jordan, and simple symmetric ordering rule. These were then used to investigate the traversal time of a relativistic quantum particle across a square barrier. We have shown that  tunneling is still instantaneous for the three ordering rules despite a relativistic treatment of time as a dynamical observable, provided that the barrier height is less than the rest mass energy. This result is similar to the earlier work of Galapon \cite{PhysRevLett.108.170402} for a non-relativistic particle. That is, tunneling is instantaneous and that only the above barrier energy components of the initial wavepacket's momentum distribution contribute to the barrier traversal time. 

The results of this paper implies that instantaneous tunneling time, or generally superluminal tunneling times, across a square barrier is not a consequence of using a non-relativistic theory but is an inherent quantum effect in the context of arrival times. However, this instantaneous tunneling can only be observed if the following conditions are satisfied: (i) the initial incident wavepacket $\psi(q)$ must be spatially wide to ensure that all the momentum components are below the barrier; and (ii) the initial incident wavepacket must be placed very far from the barrier to prevent any `leaking' into the barrier. 

It remains to be explored the case when $V_o>\mu_o c^2$, which can be done by modifying the contour in Fig. \ref{fig:contourexpecval}. By doing so, it is expected that one should be able to extract a non-zero value for the below-barrier contributions to the effective IOR of the barrier. The caveat is that the effects of spontaneous pair creation and annihilation may be significant in this regime such that the concept of TOA loses its meaning. That is, the particle that arrived may not be the same initial particle that tunneled through the barrier such that the concept of TOA-becomes ill-defined. 

It should then be enough to use a non-relativistic theory and investigate the effects of the shape of the barrier to the measured tunneling times. It is well-known that non-linear systems such as the square barrier suffers from obstructions to quantization \cite{galapon2004shouldn}. In the non-relativistic case, the correction terms to the TKF for non-linear systems, such as the square barrier, has been recently obtained \cite{pablico2022quantum}. Applying these correction terms to the non-relativistic case may lead to non-zero tunneling times. 

We leave the problem for spin-$1/2$ particles open for future studies. Earlier studies done by Bunao and Galapon \cite{bunao2015one,bunao2015relativistic}, where the TOA-operators were obtained by solving the time-energy canonical commutation relation, have shown that $\mathsf{\hat{T}_{S-1/2}} = \mathsf{\hat{T}_{S-0}} + \mathsf{\hat{T}_{E}}$ in which, $\mathsf{\hat{T}_{S-1/2}}$ and $\mathsf{\hat{T}_{S-0}}$ are the free-particle TOA-operator for spin-$1/2$ and spin-$0$ particles, respectively. Meanwhile, $\mathsf{\hat{T}_{E}}$ is an extra term which is not invariant under parity transforms, and commutes with the Dirac-Hamiltonian. The term $\mathsf{\hat{T}_{E}}$ was then thrown away because it does not contribute anything to the conjugacy relation implying $\mathsf{\hat{T}_{S-1/2}} = \mathsf{\hat{T}_{S-0}}$, but $\mathsf{\hat{T}_{E}}$ may provide for other characteristics, roles, and physical interpretations of a time observable \cite{farrales2022conjugates}. We expect the same behavior when extending the formalism to spin-$1/2$ particles, i.e., there will be an extra term to the barrier traversal time operator for spin-$1/2$ particles. However, the quantization prescription does not impose conjugacy between the Hamiltonian and TOA-operators, as such, it is possible that this extra term cannot be simply thrown away and may lead to non-zero tunneling times.

\begin{acknowledgments}
P.C.M. Flores would like to thank D.A.L. Pablico and C.D. Tica for fruitful discussions regarding the evaluation of the divergent integrals in term-by-term integration. P.C.M. Flores acknowledges the support of the Department of Science and Technology – Science Education Institute through the ASTHRDP-NSC graduate scholarship program. 
\end{acknowledgments}

\section*{Data Availability Statement}

Data sharing is not applicable to this article as no new data were created or analyzed in this study.

\appendix

\section{Non-relativistic limit of the time kernel factors}
\label{sec:TKFlimit}

For completeness, we show how the relativistic TKFs in Sec. \ref{sec:PosSpaceRep} reduces to the known TKFs of the non-relativistic TOA operator constructed by Galapon and Magadan \cite{galapon2018quantizations}. We first evaluate the modified Weyl-ordered relativistic TKF operator as follows 
\begin{align}
	\lim_{c\rightarrow\infty}& T^{\{W\}}(q,q') \nonumber \\
	=& \dfrac{1}{2} \int_0^{\frac{q+q'}{2}}  ds \lim_{c\rightarrow\infty}\mathsf{W}_s(q,q') \nonumber \\
	=& \dfrac{1}{2} \int_0^{\frac{q+q'}{2}}  ds \lim_{c\rightarrow\infty} \left\{ \mathsf{W}_s^{(1)}(q,q') + \dfrac{2}{\pi}\int_1^\infty dz \exp[-\dfrac{\mu_o c}{\hbar}\abs{q-q'}z] \dfrac{\sqrt{z^2-1}}{z} \mathsf{W}_{s,z}^{(2)}(q,q')  \right\}
	\label{eq:Weyllimit}
\end{align}
It can easily be seen that the second term of Eq. \eqref{eq:Weyllimit} vanishes exponentially. Meanwhile, the first term of Eq. \eqref{eq:Weyllimit} reduces into
\begin{align}
	\lim_{c\rightarrow\infty} & \mathsf{W}_s^{(1)}(q,q') \nonumber \\
	=& \int_0^\infty dy e^{-y} \oint_R \dfrac{dz}{2\pi i}  \dfrac{1}{z} \lim_{c\rightarrow\infty}  \sqrt{1 + \dfrac{z^2}{\mu_o^2 c^2}}  {_0}F_1 \left[ ; 1; \dfrac{\mu_o V_s^{(W)}(q,q')}{2\hbar^2}  \left( \left(q-q'\right) - i\hbar \dfrac{y}{z} \right)^2 \mathsf{P_W}(s,z,q,q')\right] \nonumber \\
	=& \int_0^\infty dy e^{-y} \oint_R \dfrac{dz}{2\pi i}  \dfrac{1}{z}  {_0}F_1 \left[ ; 1; \dfrac{\mu_o V_s^{(W)}(q,q')}{2\hbar^2}  \left( \left(q-q'\right) - i\hbar \dfrac{y}{z} \right)^2 \right].
	\label{eq:Weyllimit2}
\end{align}
The right-hand side of Eq. \eqref{eq:Weyllimit2} is further evaluated by taking the series representation of the hypergeometric function to perform a term-by-term integration, i.e., 
\begin{align}
	\lim_{c\rightarrow\infty} & \mathsf{W}_s^{(1)}(q,q') \nonumber \\
	=& \sum_{m=0}^\infty \dfrac{1}{(1)_m m!} \left( \dfrac{\mu V_s^{(W)}(q,q')}{2\hbar^2} \right)^m \sum_{n=0}^{2m} \binom{2m}{n} (q-q')^{2m-n} \left(-i\hbar\right)^n \int_0^\infty dy e^{-y} y^n \oint_R \dfrac{dz}{2\pi i}  \dfrac{1}{z^{n+1}} \nonumber \\
	=& \sum_{m=0}^\infty \dfrac{1}{(1)_m m!} \left( \dfrac{\mu V_s^{(W)}(q,q')}{2\hbar^2} (q-q')^2\right)^m \nonumber \\
	=&{_0}F_1 \left[ ; 1; \dfrac{\mu V_s^{(W)}(q,q')}{2\hbar^2} (q-q')^2 \right]
\end{align}
Thus, we now have the non-relativistic limit of the Weyl-ordered TKF given by
\begin{equation}
	\lim_{c\rightarrow\infty} T^{\{W\}}(q,q') = \dfrac{1}{2} \int_0^{\frac{q+q'}{2}}  ds {_0}F_1 \left[ ; 1; \dfrac{\mu V_s^{(W)}(q,q')}{2\hbar^2} (q-q')^2 \right]
\end{equation}
The same process is applied to obtain the non-relativistic limit of the Born-Jordan and simple-symmetric ordered TKFs. 

\section{Further details on the evaluation of the complex-valued IOR $R_c^*$}
\label{sec:IORdetails}

Here, we provide the details on the evaluation of the contour integrals Eq. \eqref{eq:contour}. Let us first consider the following integral 
\begin{equation}
	\oint dz p(z) \left( \sqrt{1 + \dfrac{\hbar^2 z^2}{\mu_o^2 c^2}} \right)^{n+1} \left( z^2 + \dfrac{V_o^2}{\hbar^2 c^2} \right)^{-n-\frac{1}{2}},
	\label{eq:contourA} 
\end{equation}
which is separately evaluated using the left and right box contours in Fig. \ref{fig:contourexpecval}(a). It is straightforward to show that the integral Eq. \eqref{eq:contourA} will vanish along the paths $z=\pm r + i y$ since $p(z)=\abs{\tilde{\psi}(z)}^2$ vanishes as $r\rightarrow\infty$. Moreover, Eq. \eqref{eq:contourA} also vanish along the semicircular paths around the branch points $z=\delta e^{i \theta} + i (\mu_o c / \hbar)$ and $z=\delta e^{i \theta} + i (V_o / \hbar c)$ as $\delta\rightarrow 0$. Taking the difference of the non-vanishing terms of the right and left box contours will then yield
\begin{align}
	\int_{-\infty}^\infty dx &  p(x+i\epsilon) \text{csgn}(x+i\epsilon)  \left( \sqrt{1 + \dfrac{\hbar^2 (x+i\epsilon)^2}{\mu_o^2 c^2}} \right)^{n+1} \left( (x+i\epsilon)^2  +  \dfrac{V_o^2}{\hbar^2 c^2}\right)^{-n-\frac{1}{2}} \nonumber \\ 
	=& \int_0^\infty dx \left( p(x) - p(-x) \right) \left( \sqrt{1 + \dfrac{\hbar^2 x^2}{\mu_o^2 c^2}} \right)^{n+1} \left( x^2 + \dfrac{V_o^2}{\hbar^2 c^2} \right)^{-n-\frac{1}{2}} \nonumber \\
	&-i \left( 1 - (-1)^{n+1} \right) \left( -i \dfrac{\hbar^2}{\mu^2 c^2} \right)^n  \int_1^\infty dy p\left( i \dfrac{\mu c}{\hbar} y\right) \sqrt{\dfrac{y^2 -1}{y^2-\frac{V_o^2}{\mu^2 c^4}}} \left( \dfrac{\sqrt{y^2-1}}{y^2-\frac{V_o^2}{\mu^2 c^4}} \right)^n
	\label{eq:contourAfin}
\end{align}
We can similarly evaluate the integral 
\begin{align}
	\oint dz p(z)\dfrac{\frac{\hbar^2}{\mu^2 c^2}z}{y^2 + \frac{\hbar^2}{\mu^2 c^2} z^2}
	\label{eq:contourB}
\end{align}
using the contour in Fig. \ref{fig:contourexpecval}(b). It is also straightforward to show that the integral Eq. \eqref{eq:contourB} will vanish along the paths $z=\pm r + i y$ since $p(z)=\abs{\tilde{\psi}(z)}^2$ vanishes as $r\rightarrow\infty$. Using the residue theorem, it is easy to show that 
\begin{align}
	\int_{-\infty}^{\infty} dx & p(x+i\epsilon) \dfrac{ \frac{\hbar^2}{\mu^2 c^2} (x+i\epsilon)}{y^2 + \frac{\hbar^2}{\mu^2 c^2}(x+i\epsilon)^2} = \int_{-\infty}^{\infty} dx  p(x) \dfrac{ \frac{\hbar^2}{\mu^2 c^2} x}{y^2 + \frac{\hbar^2}{\mu^2 c^2}x^2} - \pi i p\left( i \dfrac{\mu c}{\hbar} y\right)
	\label{eq:contourBfin}
\end{align}
We then substitute both Eqs. \eqref{eq:contourAfin} and \eqref{eq:contourBfin} into Eq. \eqref{eq:Rc*expandA} which yields
\begin{align}
	\int_0^\infty d\zeta & \mathsf{F}_B(-V_0,\zeta) \int_{-\infty}^\infty dk e^{i k \zeta} \abs{\tilde{\psi}(k)}^2 \nonumber \\
	=& i \dfrac{\hbar}{\mu c} \int_{0}^\infty  dk  \left( \abs{\tilde{\psi}(k)}^2 - \abs{\tilde{\psi}(- k)}^2\right)  \sqrt{\dfrac{\tilde{E}_k^2}{(\tilde{E}_k-V_o)^2-\mu_o^2 c^4}}
	\label{eq:term1}
\end{align}
Last, we combine Eqs. \eqref{eq:term1} and \eqref{eq:Rc*fin} to obtain
\begin{align}
	R_c^* =& i \dfrac{\hbar k_o}{\mu c} \int_{0}^\infty  dk  \left( \abs{\tilde{\psi}(k)}^2 - \abs{\tilde{\psi}(- k)}^2\right)  \sqrt{\dfrac{\tilde{E}_k^2}{(\tilde{E}_k-V_o)^2-\mu_o^2 c^4}} \nonumber \\
	&+ k_o \dfrac{2}{\pi} \int_1^\infty dy \dfrac{\sqrt{y^2-1}}{y} \mathsf{G_B}(V_o,y) \int_{-\infty}^\infty dk \abs{\tilde{\psi}(k)}^2 \dfrac{\frac{\mu c}{\hbar} y}{k^2 + \frac{\mu^2 c^2}{\hbar^2} y^2 }. 
	\label{eq:Rc*Fin}
\end{align}
Notice that the first term of \eqref{eq:Rc*Fin} is generally complex-valued while the second term is always real-valued.

\bibliography{reltunnel.bib}

\begin{thebibliography}{79}%
\makeatletter
\providecommand \@ifxundefined [1]{%
 \@ifx{#1\undefined}
}%
\providecommand \@ifnum [1]{%
 \ifnum #1\expandafter \@firstoftwo
 \else \expandafter \@secondoftwo
 \fi
}%
\providecommand \@ifx [1]{%
 \ifx #1\expandafter \@firstoftwo
 \else \expandafter \@secondoftwo
 \fi
}%
\providecommand \natexlab [1]{#1}%
\providecommand \enquote  [1]{``#1''}%
\providecommand \bibnamefont  [1]{#1}%
\providecommand \bibfnamefont [1]{#1}%
\providecommand \citenamefont [1]{#1}%
\providecommand \href@noop [0]{\@secondoftwo}%
\providecommand \href [0]{\begingroup \@sanitize@url \@href}%
\providecommand \@href[1]{\@@startlink{#1}\@@href}%
\providecommand \@@href[1]{\endgroup#1\@@endlink}%
\providecommand \@sanitize@url [0]{\catcode `\\12\catcode `\$12\catcode
  `\&12\catcode `\#12\catcode `\^12\catcode `\_12\catcode `\%12\relax}%
\providecommand \@@startlink[1]{}%
\providecommand \@@endlink[0]{}%
\providecommand \url  [0]{\begingroup\@sanitize@url \@url }%
\providecommand \@url [1]{\endgroup\@href {#1}{\urlprefix }}%
\providecommand \urlprefix  [0]{URL }%
\providecommand \Eprint [0]{\href }%
\providecommand \doibase [0]{https://doi.org/}%
\providecommand \selectlanguage [0]{\@gobble}%
\providecommand \bibinfo  [0]{\@secondoftwo}%
\providecommand \bibfield  [0]{\@secondoftwo}%
\providecommand \translation [1]{[#1]}%
\providecommand \BibitemOpen [0]{}%
\providecommand \bibitemStop [0]{}%
\providecommand \bibitemNoStop [0]{.\EOS\space}%
\providecommand \EOS [0]{\spacefactor3000\relax}%
\providecommand \BibitemShut  [1]{\csname bibitem#1\endcsname}%
\let\auto@bib@innerbib\@empty
\bibitem [{\citenamefont {MacColl}(1932)}]{maccoll1932note}%
  \BibitemOpen
  \bibfield  {author} {\bibinfo {author} {\bibfnamefont {L.}~\bibnamefont
  {MacColl}},\ }\bibfield  {title} {\enquote {\bibinfo {title} {Note on the
  transmission and reflection of wave packets by potential barriers},}\
  }\href@noop {} {\bibfield  {journal} {\bibinfo  {journal} {Physical Review}\
  }\textbf {\bibinfo {volume} {40}},\ \bibinfo {pages} {621} (\bibinfo {year}
  {1932})}\BibitemShut {NoStop}%
\bibitem [{\citenamefont {Hartman}(1962)}]{hartman1962tunneling}%
  \BibitemOpen
  \bibfield  {author} {\bibinfo {author} {\bibfnamefont {T.~E.}\ \bibnamefont
  {Hartman}},\ }\bibfield  {title} {\enquote {\bibinfo {title} {Tunneling of a
  wave packet},}\ }\href@noop {} {\bibfield  {journal} {\bibinfo  {journal}
  {Journal of Applied Physics}\ }\textbf {\bibinfo {volume} {33}},\ \bibinfo
  {pages} {3427--3433} (\bibinfo {year} {1962})}\BibitemShut {NoStop}%
\bibitem [{\citenamefont {Hauge}\ and\ \citenamefont
  {St{\o}vneng}(1989)}]{hauge1989tunneling}%
  \BibitemOpen
  \bibfield  {author} {\bibinfo {author} {\bibfnamefont {E.}~\bibnamefont
  {Hauge}}\ and\ \bibinfo {author} {\bibfnamefont {J.}~\bibnamefont
  {St{\o}vneng}},\ }\bibfield  {title} {\enquote {\bibinfo {title} {Tunneling
  times: a critical review},}\ }\href@noop {} {\bibfield  {journal} {\bibinfo
  {journal} {Reviews of Modern Physics}\ }\textbf {\bibinfo {volume} {61}},\
  \bibinfo {pages} {917} (\bibinfo {year} {1989})}\BibitemShut {NoStop}%
\bibitem [{\citenamefont {Landauer}\ and\ \citenamefont
  {Martin}(1994)}]{landauer1994barrier}%
  \BibitemOpen
  \bibfield  {author} {\bibinfo {author} {\bibfnamefont {R.}~\bibnamefont
  {Landauer}}\ and\ \bibinfo {author} {\bibfnamefont {T.}~\bibnamefont
  {Martin}},\ }\bibfield  {title} {\enquote {\bibinfo {title} {Barrier
  interaction time in tunneling},}\ }\href@noop {} {\bibfield  {journal}
  {\bibinfo  {journal} {Reviews of Modern Physics}\ }\textbf {\bibinfo {volume}
  {66}},\ \bibinfo {pages} {217} (\bibinfo {year} {1994})}\BibitemShut
  {NoStop}%
\bibitem [{\citenamefont {Pauli}\ \emph {et~al.}(1933)\citenamefont {Pauli}
  \emph {et~al.}}]{pauli1933handbuch}%
  \BibitemOpen
  \bibfield  {author} {\bibinfo {author} {\bibfnamefont {W.}~\bibnamefont
  {Pauli}} \emph {et~al.},\ }\bibfield  {title} {\enquote {\bibinfo {title}
  {Handbuch der physik},}\ }\href@noop {} {\bibfield  {journal} {\bibinfo
  {journal} {Geiger and scheel}\ }\textbf {\bibinfo {volume} {2}},\ \bibinfo
  {pages} {83--272} (\bibinfo {year} {1933})}\BibitemShut {NoStop}%
\bibitem [{\citenamefont {Wigner}(1955)}]{wigner1955lower}%
  \BibitemOpen
  \bibfield  {author} {\bibinfo {author} {\bibfnamefont {E.~P.}\ \bibnamefont
  {Wigner}},\ }\bibfield  {title} {\enquote {\bibinfo {title} {Lower limit for
  the energy derivative of the scattering phase shift},}\ }\href@noop {}
  {\bibfield  {journal} {\bibinfo  {journal} {Physical Review}\ }\textbf
  {\bibinfo {volume} {98}},\ \bibinfo {pages} {145} (\bibinfo {year}
  {1955})}\BibitemShut {NoStop}%
\bibitem [{\citenamefont {B{\"u}ttiker}\ and\ \citenamefont
  {Landauer}(1982)}]{buttiker1982traversal}%
  \BibitemOpen
  \bibfield  {author} {\bibinfo {author} {\bibfnamefont {M.}~\bibnamefont
  {B{\"u}ttiker}}\ and\ \bibinfo {author} {\bibfnamefont {R.}~\bibnamefont
  {Landauer}},\ }\bibfield  {title} {\enquote {\bibinfo {title} {Traversal time
  for tunneling},}\ }\href@noop {} {\bibfield  {journal} {\bibinfo  {journal}
  {Physical Review Letters}\ }\textbf {\bibinfo {volume} {49}},\ \bibinfo
  {pages} {1739} (\bibinfo {year} {1982})}\BibitemShut {NoStop}%
\bibitem [{\citenamefont {Baz}(1966)}]{baz1966lifetime}%
  \BibitemOpen
  \bibfield  {author} {\bibinfo {author} {\bibfnamefont {A.}~\bibnamefont
  {Baz}},\ }\bibfield  {title} {\enquote {\bibinfo {title} {Lifetime of
  intermediate states},}\ }\href@noop {} {\bibfield  {journal} {\bibinfo
  {journal} {Yadern. Fiz.}\ }\textbf {\bibinfo {volume} {4}} (\bibinfo {year}
  {1966})}\BibitemShut {NoStop}%
\bibitem [{\citenamefont {Rybachenko}(1967)}]{rybachenko1967time}%
  \BibitemOpen
  \bibfield  {author} {\bibinfo {author} {\bibfnamefont {V.}~\bibnamefont
  {Rybachenko}},\ }\bibfield  {title} {\enquote {\bibinfo {title} {Time of
  penetration of a particle through a potential barrier},}\ }\href@noop {}
  {\bibfield  {journal} {\bibinfo  {journal} {Sov. J. Nucl. Phys.}\ }\textbf
  {\bibinfo {volume} {5}},\ \bibinfo {pages} {635--639} (\bibinfo {year}
  {1967})}\BibitemShut {NoStop}%
\bibitem [{\citenamefont {B{\"u}ttiker}(1983)}]{buttiker1983larmor}%
  \BibitemOpen
  \bibfield  {author} {\bibinfo {author} {\bibfnamefont {M.}~\bibnamefont
  {B{\"u}ttiker}},\ }\bibfield  {title} {\enquote {\bibinfo {title} {Larmor
  precession and the traversal time for tunneling},}\ }\href@noop {} {\bibfield
   {journal} {\bibinfo  {journal} {Physical Review B}\ }\textbf {\bibinfo
  {volume} {27}},\ \bibinfo {pages} {6178} (\bibinfo {year}
  {1983})}\BibitemShut {NoStop}%
\bibitem [{\citenamefont {Pollak}\ and\ \citenamefont
  {Miller}(1984)}]{pollak1984new}%
  \BibitemOpen
  \bibfield  {author} {\bibinfo {author} {\bibfnamefont {E.}~\bibnamefont
  {Pollak}}\ and\ \bibinfo {author} {\bibfnamefont {W.~H.}\ \bibnamefont
  {Miller}},\ }\bibfield  {title} {\enquote {\bibinfo {title} {New physical
  interpretation for time in scattering theory},}\ }\href@noop {} {\bibfield
  {journal} {\bibinfo  {journal} {Physical review letters}\ }\textbf {\bibinfo
  {volume} {53}},\ \bibinfo {pages} {115} (\bibinfo {year} {1984})}\BibitemShut
  {NoStop}%
\bibitem [{\citenamefont {Smith}(1960)}]{smith1960lifetime}%
  \BibitemOpen
  \bibfield  {author} {\bibinfo {author} {\bibfnamefont {F.~T.}\ \bibnamefont
  {Smith}},\ }\bibfield  {title} {\enquote {\bibinfo {title} {Lifetime matrix
  in collision theory},}\ }\href@noop {} {\bibfield  {journal} {\bibinfo
  {journal} {Physical Review}\ }\textbf {\bibinfo {volume} {118}},\ \bibinfo
  {pages} {349} (\bibinfo {year} {1960})}\BibitemShut {NoStop}%
\bibitem [{\citenamefont {Sokolovski}\ and\ \citenamefont
  {Baskin}(1987)}]{sokolovski1987traversal}%
  \BibitemOpen
  \bibfield  {author} {\bibinfo {author} {\bibfnamefont {D.}~\bibnamefont
  {Sokolovski}}\ and\ \bibinfo {author} {\bibfnamefont {L.}~\bibnamefont
  {Baskin}},\ }\bibfield  {title} {\enquote {\bibinfo {title} {Traversal time
  in quantum scattering},}\ }\href@noop {} {\bibfield  {journal} {\bibinfo
  {journal} {Physical Review A}\ }\textbf {\bibinfo {volume} {36}},\ \bibinfo
  {pages} {4604} (\bibinfo {year} {1987})}\BibitemShut {NoStop}%
\bibitem [{\citenamefont {Yamada}(2004)}]{yamada2004unified}%
  \BibitemOpen
  \bibfield  {author} {\bibinfo {author} {\bibfnamefont {N.}~\bibnamefont
  {Yamada}},\ }\bibfield  {title} {\enquote {\bibinfo {title} {Unified
  derivation of tunneling times from decoherence functionals},}\ }\href@noop {}
  {\bibfield  {journal} {\bibinfo  {journal} {Physical review letters}\
  }\textbf {\bibinfo {volume} {93}},\ \bibinfo {pages} {170401} (\bibinfo
  {year} {2004})}\BibitemShut {NoStop}%
\bibitem [{\citenamefont {Galapon}(2012)}]{PhysRevLett.108.170402}%
  \BibitemOpen
  \bibfield  {author} {\bibinfo {author} {\bibfnamefont {E.~A.}\ \bibnamefont
  {Galapon}},\ }\bibfield  {title} {\enquote {\bibinfo {title} {Only above
  barrier energy components contribute to barrier traversal time},}\
  }\href@noop {} {\bibfield  {journal} {\bibinfo  {journal} {Physical review
  letters}\ }\textbf {\bibinfo {volume} {108}},\ \bibinfo {pages} {170402}
  (\bibinfo {year} {2012})}\BibitemShut {NoStop}%
\bibitem [{\citenamefont {de~Carvalho}\ and\ \citenamefont
  {Nussenzveig}(2002)}]{de2002time}%
  \BibitemOpen
  \bibfield  {author} {\bibinfo {author} {\bibfnamefont {C.~A.}\ \bibnamefont
  {de~Carvalho}}\ and\ \bibinfo {author} {\bibfnamefont {H.~M.}\ \bibnamefont
  {Nussenzveig}},\ }\bibfield  {title} {\enquote {\bibinfo {title} {Time
  delay},}\ }\href@noop {} {\bibfield  {journal} {\bibinfo  {journal} {Physics
  Reports}\ }\textbf {\bibinfo {volume} {364}},\ \bibinfo {pages} {83--174}
  (\bibinfo {year} {2002})}\BibitemShut {NoStop}%
\bibitem [{\citenamefont {Winful}(2006)}]{winful2006tunneling}%
  \BibitemOpen
  \bibfield  {author} {\bibinfo {author} {\bibfnamefont {H.~G.}\ \bibnamefont
  {Winful}},\ }\bibfield  {title} {\enquote {\bibinfo {title} {Tunneling time,
  the hartman effect, and superluminality: A proposed resolution of an old
  paradox},}\ }\href@noop {} {\bibfield  {journal} {\bibinfo  {journal}
  {Physics Reports}\ }\textbf {\bibinfo {volume} {436}},\ \bibinfo {pages}
  {1--69} (\bibinfo {year} {2006})}\BibitemShut {NoStop}%
\bibitem [{\citenamefont {Imafuku}, \citenamefont {Ohba},\ and\ \citenamefont
  {Yamanaka}(1997)}]{imafuku1997effects}%
  \BibitemOpen
  \bibfield  {author} {\bibinfo {author} {\bibfnamefont {K.}~\bibnamefont
  {Imafuku}}, \bibinfo {author} {\bibfnamefont {I.}~\bibnamefont {Ohba}},\ and\
  \bibinfo {author} {\bibfnamefont {Y.}~\bibnamefont {Yamanaka}},\ }\bibfield
  {title} {\enquote {\bibinfo {title} {Effects of inelastic scattering on
  tunneling time based on the generalized diffusion process approach},}\
  }\href@noop {} {\bibfield  {journal} {\bibinfo  {journal} {Physical Review
  A}\ }\textbf {\bibinfo {volume} {56}},\ \bibinfo {pages} {1142} (\bibinfo
  {year} {1997})}\BibitemShut {NoStop}%
\bibitem [{\citenamefont {Brouard}, \citenamefont {Sala},\ and\ \citenamefont
  {Muga}(1994)}]{brouard1994systematic}%
  \BibitemOpen
  \bibfield  {author} {\bibinfo {author} {\bibfnamefont {S.}~\bibnamefont
  {Brouard}}, \bibinfo {author} {\bibfnamefont {R.}~\bibnamefont {Sala}},\ and\
  \bibinfo {author} {\bibfnamefont {J.}~\bibnamefont {Muga}},\ }\bibfield
  {title} {\enquote {\bibinfo {title} {Systematic approach to define and
  classify quantum transmission and reflection times},}\ }\href@noop {}
  {\bibfield  {journal} {\bibinfo  {journal} {Physical Review A}\ }\textbf
  {\bibinfo {volume} {49}},\ \bibinfo {pages} {4312} (\bibinfo {year}
  {1994})}\BibitemShut {NoStop}%
\bibitem [{\citenamefont {Jaworski}\ and\ \citenamefont
  {Wardlaw}(1988)}]{jaworski1988time}%
  \BibitemOpen
  \bibfield  {author} {\bibinfo {author} {\bibfnamefont {W.}~\bibnamefont
  {Jaworski}}\ and\ \bibinfo {author} {\bibfnamefont {D.~M.}\ \bibnamefont
  {Wardlaw}},\ }\bibfield  {title} {\enquote {\bibinfo {title} {Time delay in
  tunneling: Sojourn-time approach versus mean-position approach},}\
  }\href@noop {} {\bibfield  {journal} {\bibinfo  {journal} {Physical Review
  A}\ }\textbf {\bibinfo {volume} {38}},\ \bibinfo {pages} {5404} (\bibinfo
  {year} {1988})}\BibitemShut {NoStop}%
\bibitem [{\citenamefont {Leavens}\ and\ \citenamefont
  {Aers}(1989)}]{leavens1989dwell}%
  \BibitemOpen
  \bibfield  {author} {\bibinfo {author} {\bibfnamefont {C.}~\bibnamefont
  {Leavens}}\ and\ \bibinfo {author} {\bibfnamefont {G.}~\bibnamefont {Aers}},\
  }\bibfield  {title} {\enquote {\bibinfo {title} {Dwell time and phase times
  for transmission and reflection},}\ }\href@noop {} {\bibfield  {journal}
  {\bibinfo  {journal} {Physical Review B}\ }\textbf {\bibinfo {volume} {39}},\
  \bibinfo {pages} {1202} (\bibinfo {year} {1989})}\BibitemShut {NoStop}%
\bibitem [{\citenamefont {Hauge}, \citenamefont {Falck},\ and\ \citenamefont
  {Fjeldly}(1987)}]{hauge1987transmission}%
  \BibitemOpen
  \bibfield  {author} {\bibinfo {author} {\bibfnamefont {E.}~\bibnamefont
  {Hauge}}, \bibinfo {author} {\bibfnamefont {J.}~\bibnamefont {Falck}},\ and\
  \bibinfo {author} {\bibfnamefont {T.}~\bibnamefont {Fjeldly}},\ }\bibfield
  {title} {\enquote {\bibinfo {title} {Transmission and reflection times for
  scattering of wave packets off tunneling barriers},}\ }\href@noop {}
  {\bibfield  {journal} {\bibinfo  {journal} {Physical Review B}\ }\textbf
  {\bibinfo {volume} {36}},\ \bibinfo {pages} {4203} (\bibinfo {year}
  {1987})}\BibitemShut {NoStop}%
\bibitem [{\citenamefont {Galapon}(2002)}]{galapon2002pauli}%
  \BibitemOpen
  \bibfield  {author} {\bibinfo {author} {\bibfnamefont {E.}~\bibnamefont
  {Galapon}},\ }\bibfield  {title} {\enquote {\bibinfo {title} {Pauli's theorem
  and quantum canonical pairs: the consistency of a bounded, self--adjoint time
  operator canonically conjugate to a hamiltonian with non--empty point
  spectrum},}\ }\href@noop {} {\bibfield  {journal} {\bibinfo  {journal}
  {Proceedings of the Royal Society of London. Series A: Mathematical, Physical
  and Engineering Sciences}\ }\textbf {\bibinfo {volume} {458}},\ \bibinfo
  {pages} {451--472} (\bibinfo {year} {2002})}\BibitemShut {NoStop}%
\bibitem [{\citenamefont {Winful}(2003)}]{winful2003nature}%
  \BibitemOpen
  \bibfield  {author} {\bibinfo {author} {\bibfnamefont {H.~G.}\ \bibnamefont
  {Winful}},\ }\bibfield  {title} {\enquote {\bibinfo {title} {Nature of
  “superluminal" barrier tunneling},}\ }\href@noop {} {\bibfield  {journal}
  {\bibinfo  {journal} {Physical review letters}\ }\textbf {\bibinfo {volume}
  {90}},\ \bibinfo {pages} {023901} (\bibinfo {year} {2003})}\BibitemShut
  {NoStop}%
\bibitem [{\citenamefont {Eckle}\ \emph
  {et~al.}(2008{\natexlab{a}})\citenamefont {Eckle}, \citenamefont {Smolarski},
  \citenamefont {Schlup}, \citenamefont {Biegert}, \citenamefont {Staudte},
  \citenamefont {Sch{\"o}ffler}, \citenamefont {Muller}, \citenamefont
  {D{\"o}rner},\ and\ \citenamefont
  {Keller}}]{eckleAttosecondAngularStreaking2008}%
  \BibitemOpen
  \bibfield  {author} {\bibinfo {author} {\bibfnamefont {P.}~\bibnamefont
  {Eckle}}, \bibinfo {author} {\bibfnamefont {M.}~\bibnamefont {Smolarski}},
  \bibinfo {author} {\bibfnamefont {P.}~\bibnamefont {Schlup}}, \bibinfo
  {author} {\bibfnamefont {J.}~\bibnamefont {Biegert}}, \bibinfo {author}
  {\bibfnamefont {A.}~\bibnamefont {Staudte}}, \bibinfo {author} {\bibfnamefont
  {M.}~\bibnamefont {Sch{\"o}ffler}}, \bibinfo {author} {\bibfnamefont {H.~G.}\
  \bibnamefont {Muller}}, \bibinfo {author} {\bibfnamefont {R.}~\bibnamefont
  {D{\"o}rner}},\ and\ \bibinfo {author} {\bibfnamefont {U.}~\bibnamefont
  {Keller}},\ }\bibfield  {title} {\enquote {\bibinfo {title} {Attosecond
  angular streaking},}\ }\href {https://doi.org/10.1038/nphys982} {\bibfield
  {journal} {\bibinfo  {journal} {Nature Physics}\ }\textbf {\bibinfo {volume}
  {4}},\ \bibinfo {pages} {565--570} (\bibinfo {year}
  {2008}{\natexlab{a}})}\BibitemShut {NoStop}%
\bibitem [{\citenamefont {Eckle}\ \emph
  {et~al.}(2008{\natexlab{b}})\citenamefont {Eckle}, \citenamefont {Pfeiffer},
  \citenamefont {Cirelli}, \citenamefont {Staudte}, \citenamefont {Dörner},
  \citenamefont {Muller}, \citenamefont {Büttiker},\ and\ \citenamefont
  {Keller}}]{doi:10.1126/science.1163439}%
  \BibitemOpen
  \bibfield  {author} {\bibinfo {author} {\bibfnamefont {P.}~\bibnamefont
  {Eckle}}, \bibinfo {author} {\bibfnamefont {A.~N.}\ \bibnamefont {Pfeiffer}},
  \bibinfo {author} {\bibfnamefont {C.}~\bibnamefont {Cirelli}}, \bibinfo
  {author} {\bibfnamefont {A.}~\bibnamefont {Staudte}}, \bibinfo {author}
  {\bibfnamefont {R.}~\bibnamefont {Dörner}}, \bibinfo {author} {\bibfnamefont
  {H.~G.}\ \bibnamefont {Muller}}, \bibinfo {author} {\bibfnamefont
  {M.}~\bibnamefont {Büttiker}},\ and\ \bibinfo {author} {\bibfnamefont
  {U.}~\bibnamefont {Keller}},\ }\bibfield  {title} {\enquote {\bibinfo {title}
  {Attosecond ionization and tunneling delay time measurements in helium},}\
  }\href {https://doi.org/10.1126/science.1163439} {\bibfield  {journal}
  {\bibinfo  {journal} {Science}\ }\textbf {\bibinfo {volume} {322}},\ \bibinfo
  {pages} {1525--1529} (\bibinfo {year} {2008}{\natexlab{b}})}\BibitemShut
  {NoStop}%
\bibitem [{\citenamefont {Pfeiffer}\ \emph {et~al.}(2012)\citenamefont
  {Pfeiffer}, \citenamefont {Cirelli}, \citenamefont {Smolarski}, \citenamefont
  {Dimitrovski}, \citenamefont {Abu-samha}, \citenamefont {Madsen},\ and\
  \citenamefont {Keller}}]{pfeifferAttoclockRevealsNatural2012}%
  \BibitemOpen
  \bibfield  {author} {\bibinfo {author} {\bibfnamefont {A.~N.}\ \bibnamefont
  {Pfeiffer}}, \bibinfo {author} {\bibfnamefont {C.}~\bibnamefont {Cirelli}},
  \bibinfo {author} {\bibfnamefont {M.}~\bibnamefont {Smolarski}}, \bibinfo
  {author} {\bibfnamefont {D.}~\bibnamefont {Dimitrovski}}, \bibinfo {author}
  {\bibfnamefont {M.}~\bibnamefont {Abu-samha}}, \bibinfo {author}
  {\bibfnamefont {L.~B.}\ \bibnamefont {Madsen}},\ and\ \bibinfo {author}
  {\bibfnamefont {U.}~\bibnamefont {Keller}},\ }\bibfield  {title} {\enquote
  {\bibinfo {title} {Attoclock reveals natural coordinates of the laser-induced
  tunnelling current flow in atoms},}\ }\href
  {https://doi.org/10.1038/nphys2125} {\bibfield  {journal} {\bibinfo
  {journal} {Nature Physics}\ }\textbf {\bibinfo {volume} {8}},\ \bibinfo
  {pages} {76--80} (\bibinfo {year} {2012})}\BibitemShut {NoStop}%
\bibitem [{\citenamefont {Pfeiffer}\ \emph {et~al.}(2013)\citenamefont
  {Pfeiffer}, \citenamefont {Cirelli}, \citenamefont {Smolarski},\ and\
  \citenamefont {Keller}}]{pfeiffer2013recent}%
  \BibitemOpen
  \bibfield  {author} {\bibinfo {author} {\bibfnamefont {A.~N.}\ \bibnamefont
  {Pfeiffer}}, \bibinfo {author} {\bibfnamefont {C.}~\bibnamefont {Cirelli}},
  \bibinfo {author} {\bibfnamefont {M.}~\bibnamefont {Smolarski}},\ and\
  \bibinfo {author} {\bibfnamefont {U.}~\bibnamefont {Keller}},\ }\bibfield
  {title} {\enquote {\bibinfo {title} {Recent attoclock measurements of strong
  field ionization},}\ }\href@noop {} {\bibfield  {journal} {\bibinfo
  {journal} {Chemical Physics}\ }\textbf {\bibinfo {volume} {414}},\ \bibinfo
  {pages} {84--91} (\bibinfo {year} {2013})}\BibitemShut {NoStop}%
\bibitem [{\citenamefont {Sainadh}\ \emph {et~al.}(2019)\citenamefont
  {Sainadh}, \citenamefont {Xu}, \citenamefont {Wang}, \citenamefont
  {Atia-Tul-Noor}, \citenamefont {Wallace}, \citenamefont {Douguet},
  \citenamefont {Bray}, \citenamefont {Ivanov}, \citenamefont {Bartschat},
  \citenamefont {Kheifets} \emph {et~al.}}]{sainadh2019attosecond}%
  \BibitemOpen
  \bibfield  {author} {\bibinfo {author} {\bibfnamefont {U.~S.}\ \bibnamefont
  {Sainadh}}, \bibinfo {author} {\bibfnamefont {H.}~\bibnamefont {Xu}},
  \bibinfo {author} {\bibfnamefont {X.}~\bibnamefont {Wang}}, \bibinfo {author}
  {\bibfnamefont {A.}~\bibnamefont {Atia-Tul-Noor}}, \bibinfo {author}
  {\bibfnamefont {W.~C.}\ \bibnamefont {Wallace}}, \bibinfo {author}
  {\bibfnamefont {N.}~\bibnamefont {Douguet}}, \bibinfo {author} {\bibfnamefont
  {A.}~\bibnamefont {Bray}}, \bibinfo {author} {\bibfnamefont {I.}~\bibnamefont
  {Ivanov}}, \bibinfo {author} {\bibfnamefont {K.}~\bibnamefont {Bartschat}},
  \bibinfo {author} {\bibfnamefont {A.}~\bibnamefont {Kheifets}}, \emph
  {et~al.},\ }\bibfield  {title} {\enquote {\bibinfo {title} {Attosecond
  angular streaking and tunnelling time in atomic hydrogen},}\ }\href@noop {}
  {\bibfield  {journal} {\bibinfo  {journal} {Nature}\ }\textbf {\bibinfo
  {volume} {568}},\ \bibinfo {pages} {75--77} (\bibinfo {year}
  {2019})}\BibitemShut {NoStop}%
\bibitem [{\citenamefont {Torlina}\ \emph {et~al.}(2015)\citenamefont
  {Torlina}, \citenamefont {Morales}, \citenamefont {Kaushal}, \citenamefont
  {Ivanov}, \citenamefont {Kheifets}, \citenamefont {Zielinski}, \citenamefont
  {Scrinzi}, \citenamefont {Muller}, \citenamefont {Sukiasyan}, \citenamefont
  {Ivanov} \emph {et~al.}}]{torlina2015interpreting}%
  \BibitemOpen
  \bibfield  {author} {\bibinfo {author} {\bibfnamefont {L.}~\bibnamefont
  {Torlina}}, \bibinfo {author} {\bibfnamefont {F.}~\bibnamefont {Morales}},
  \bibinfo {author} {\bibfnamefont {J.}~\bibnamefont {Kaushal}}, \bibinfo
  {author} {\bibfnamefont {I.}~\bibnamefont {Ivanov}}, \bibinfo {author}
  {\bibfnamefont {A.}~\bibnamefont {Kheifets}}, \bibinfo {author}
  {\bibfnamefont {A.}~\bibnamefont {Zielinski}}, \bibinfo {author}
  {\bibfnamefont {A.}~\bibnamefont {Scrinzi}}, \bibinfo {author} {\bibfnamefont
  {H.~G.}\ \bibnamefont {Muller}}, \bibinfo {author} {\bibfnamefont
  {S.}~\bibnamefont {Sukiasyan}}, \bibinfo {author} {\bibfnamefont
  {M.}~\bibnamefont {Ivanov}}, \emph {et~al.},\ }\bibfield  {title} {\enquote
  {\bibinfo {title} {Interpreting attoclock measurements of tunnelling
  times},}\ }\href@noop {} {\bibfield  {journal} {\bibinfo  {journal} {Nature
  Physics}\ }\textbf {\bibinfo {volume} {11}},\ \bibinfo {pages} {503--508}
  (\bibinfo {year} {2015})}\BibitemShut {NoStop}%
\bibitem [{\citenamefont {Landsman}\ \emph {et~al.}(2014)\citenamefont
  {Landsman}, \citenamefont {Weger}, \citenamefont {Maurer}, \citenamefont
  {Boge}, \citenamefont {Ludwig}, \citenamefont {Heuser}, \citenamefont
  {Cirelli}, \citenamefont {Gallmann},\ and\ \citenamefont
  {Keller}}]{landsman2014ultrafast}%
  \BibitemOpen
  \bibfield  {author} {\bibinfo {author} {\bibfnamefont {A.~S.}\ \bibnamefont
  {Landsman}}, \bibinfo {author} {\bibfnamefont {M.}~\bibnamefont {Weger}},
  \bibinfo {author} {\bibfnamefont {J.}~\bibnamefont {Maurer}}, \bibinfo
  {author} {\bibfnamefont {R.}~\bibnamefont {Boge}}, \bibinfo {author}
  {\bibfnamefont {A.}~\bibnamefont {Ludwig}}, \bibinfo {author} {\bibfnamefont
  {S.}~\bibnamefont {Heuser}}, \bibinfo {author} {\bibfnamefont
  {C.}~\bibnamefont {Cirelli}}, \bibinfo {author} {\bibfnamefont
  {L.}~\bibnamefont {Gallmann}},\ and\ \bibinfo {author} {\bibfnamefont
  {U.}~\bibnamefont {Keller}},\ }\bibfield  {title} {\enquote {\bibinfo {title}
  {Ultrafast resolution of tunneling delay time},}\ }\href@noop {} {\bibfield
  {journal} {\bibinfo  {journal} {Optica}\ }\textbf {\bibinfo {volume} {1}},\
  \bibinfo {pages} {343--349} (\bibinfo {year} {2014})}\BibitemShut {NoStop}%
\bibitem [{\citenamefont {Camus}\ \emph {et~al.}(2017)\citenamefont {Camus},
  \citenamefont {Yakaboylu}, \citenamefont {Fechner}, \citenamefont {Klaiber},
  \citenamefont {Laux}, \citenamefont {Mi}, \citenamefont {Hatsagortsyan},
  \citenamefont {Pfeifer}, \citenamefont {Keitel},\ and\ \citenamefont
  {Moshammer}}]{camus2017experimental}%
  \BibitemOpen
  \bibfield  {author} {\bibinfo {author} {\bibfnamefont {N.}~\bibnamefont
  {Camus}}, \bibinfo {author} {\bibfnamefont {E.}~\bibnamefont {Yakaboylu}},
  \bibinfo {author} {\bibfnamefont {L.}~\bibnamefont {Fechner}}, \bibinfo
  {author} {\bibfnamefont {M.}~\bibnamefont {Klaiber}}, \bibinfo {author}
  {\bibfnamefont {M.}~\bibnamefont {Laux}}, \bibinfo {author} {\bibfnamefont
  {Y.}~\bibnamefont {Mi}}, \bibinfo {author} {\bibfnamefont {K.~Z.}\
  \bibnamefont {Hatsagortsyan}}, \bibinfo {author} {\bibfnamefont
  {T.}~\bibnamefont {Pfeifer}}, \bibinfo {author} {\bibfnamefont {C.~H.}\
  \bibnamefont {Keitel}},\ and\ \bibinfo {author} {\bibfnamefont
  {R.}~\bibnamefont {Moshammer}},\ }\bibfield  {title} {\enquote {\bibinfo
  {title} {Experimental evidence for quantum tunneling time},}\ }\href@noop {}
  {\bibfield  {journal} {\bibinfo  {journal} {Physical review letters}\
  }\textbf {\bibinfo {volume} {119}},\ \bibinfo {pages} {023201} (\bibinfo
  {year} {2017})}\BibitemShut {NoStop}%
\bibitem [{\citenamefont {Ramos}\ \emph {et~al.}(2020)\citenamefont {Ramos},
  \citenamefont {Spierings}, \citenamefont {Racicot},\ and\ \citenamefont
  {Steinberg}}]{ramos2020measurement}%
  \BibitemOpen
  \bibfield  {author} {\bibinfo {author} {\bibfnamefont {R.}~\bibnamefont
  {Ramos}}, \bibinfo {author} {\bibfnamefont {D.}~\bibnamefont {Spierings}},
  \bibinfo {author} {\bibfnamefont {I.}~\bibnamefont {Racicot}},\ and\ \bibinfo
  {author} {\bibfnamefont {A.~M.}\ \bibnamefont {Steinberg}},\ }\bibfield
  {title} {\enquote {\bibinfo {title} {Measurement of the time spent by a
  tunnelling atom within the barrier region},}\ }\href@noop {} {\bibfield
  {journal} {\bibinfo  {journal} {Nature}\ }\textbf {\bibinfo {volume} {583}},\
  \bibinfo {pages} {529--532} (\bibinfo {year} {2020})}\BibitemShut {NoStop}%
\bibitem [{\citenamefont {Spierings}\ and\ \citenamefont
  {Steinberg}(2021)}]{PhysRevLett.127.133001}%
  \BibitemOpen
  \bibfield  {author} {\bibinfo {author} {\bibfnamefont {D.~C.}\ \bibnamefont
  {Spierings}}\ and\ \bibinfo {author} {\bibfnamefont {A.~M.}\ \bibnamefont
  {Steinberg}},\ }\bibfield  {title} {\enquote {\bibinfo {title} {Observation
  of the decrease of larmor tunneling times with lower incident energy},}\
  }\href {https://doi.org/10.1103/PhysRevLett.127.133001} {\bibfield  {journal}
  {\bibinfo  {journal} {Phys. Rev. Lett.}\ }\textbf {\bibinfo {volume} {127}},\
  \bibinfo {pages} {133001} (\bibinfo {year} {2021})}\BibitemShut {NoStop}%
\bibitem [{\citenamefont {De~Leo}\ and\ \citenamefont
  {Rotelli}(2007)}]{de2007dirac}%
  \BibitemOpen
  \bibfield  {author} {\bibinfo {author} {\bibfnamefont {S.}~\bibnamefont
  {De~Leo}}\ and\ \bibinfo {author} {\bibfnamefont {P.~P.}\ \bibnamefont
  {Rotelli}},\ }\bibfield  {title} {\enquote {\bibinfo {title} {Dirac equation
  studies in the tunneling energy zone},}\ }\href@noop {} {\bibfield  {journal}
  {\bibinfo  {journal} {The European Physical Journal C}\ }\textbf {\bibinfo
  {volume} {51}},\ \bibinfo {pages} {241--247} (\bibinfo {year}
  {2007})}\BibitemShut {NoStop}%
\bibitem [{\citenamefont {De~Leo}(2013)}]{de2013study}%
  \BibitemOpen
  \bibfield  {author} {\bibinfo {author} {\bibfnamefont {S.}~\bibnamefont
  {De~Leo}},\ }\bibfield  {title} {\enquote {\bibinfo {title} {A study of
  transit times in dirac tunneling},}\ }\href@noop {} {\bibfield  {journal}
  {\bibinfo  {journal} {Journal of Physics A: Mathematical and Theoretical}\
  }\textbf {\bibinfo {volume} {46}},\ \bibinfo {pages} {155306} (\bibinfo
  {year} {2013})}\BibitemShut {NoStop}%
\bibitem [{\citenamefont {Petrillo}\ and\ \citenamefont
  {Janner}(2003)}]{PhysRevA.67.012110}%
  \BibitemOpen
  \bibfield  {author} {\bibinfo {author} {\bibfnamefont {V.}~\bibnamefont
  {Petrillo}}\ and\ \bibinfo {author} {\bibfnamefont {D.}~\bibnamefont
  {Janner}},\ }\bibfield  {title} {\enquote {\bibinfo {title} {Relativistic
  analysis of a wave packet interacting with a quantum-mechanical barrier},}\
  }\href {https://doi.org/10.1103/PhysRevA.67.012110} {\bibfield  {journal}
  {\bibinfo  {journal} {Phys. Rev. A}\ }\textbf {\bibinfo {volume} {67}},\
  \bibinfo {pages} {012110} (\bibinfo {year} {2003})}\BibitemShut {NoStop}%
\bibitem [{\citenamefont {Krekora}, \citenamefont {Su},\ and\ \citenamefont
  {Grobe}(2001)}]{krekora2001effects}%
  \BibitemOpen
  \bibfield  {author} {\bibinfo {author} {\bibfnamefont {P.}~\bibnamefont
  {Krekora}}, \bibinfo {author} {\bibfnamefont {Q.}~\bibnamefont {Su}},\ and\
  \bibinfo {author} {\bibfnamefont {R.}~\bibnamefont {Grobe}},\ }\bibfield
  {title} {\enquote {\bibinfo {title} {Effects of relativity on the
  time-resolved tunneling of electron wave packets},}\ }\href@noop {}
  {\bibfield  {journal} {\bibinfo  {journal} {Physical Review A}\ }\textbf
  {\bibinfo {volume} {63}},\ \bibinfo {pages} {032107} (\bibinfo {year}
  {2001})}\BibitemShut {NoStop}%
\bibitem [{\citenamefont {Flores}\ and\ \citenamefont
  {Galapon}(2023)}]{flores2022letter}%
  \BibitemOpen
  \bibfield  {author} {\bibinfo {author} {\bibfnamefont {P.~C.}\ \bibnamefont
  {Flores}}\ and\ \bibinfo {author} {\bibfnamefont {E.~A.}\ \bibnamefont
  {Galapon}},\ }\bibfield  {title} {\enquote {\bibinfo {title} {Instantaneous
  tunneling of relativistic massive spin-0 particles},}\ }\href@noop {}
  {\bibfield  {journal} {\bibinfo  {journal} {Europhysics Letters}\ }\textbf
  {\bibinfo {volume} {141}},\ \bibinfo {pages} {10001} (\bibinfo {year}
  {2023})}\BibitemShut {NoStop}%
\bibitem [{\citenamefont {Galapon}\ and\ \citenamefont
  {Magadan}(2018)}]{galapon2018quantizations}%
  \BibitemOpen
  \bibfield  {author} {\bibinfo {author} {\bibfnamefont {E.~A.}\ \bibnamefont
  {Galapon}}\ and\ \bibinfo {author} {\bibfnamefont {J.~J.~P.}\ \bibnamefont
  {Magadan}},\ }\bibfield  {title} {\enquote {\bibinfo {title} {Quantizations
  of the classical time of arrival and their dynamics},}\ }\href@noop {}
  {\bibfield  {journal} {\bibinfo  {journal} {Annals of Physics}\ }\textbf
  {\bibinfo {volume} {397}},\ \bibinfo {pages} {278--302} (\bibinfo {year}
  {2018})}\BibitemShut {NoStop}%
\bibitem [{\citenamefont {Bohm}(1974)}]{bohm1974rigged}%
  \BibitemOpen
  \bibfield  {author} {\bibinfo {author} {\bibfnamefont {A.}~\bibnamefont
  {Bohm}},\ }\href@noop {} {\enquote {\bibinfo {title} {Rigged hilbert space
  and quantum mechanics},}\ }\bibinfo {type} {Tech. Rep.}\ (\bibinfo {year}
  {1974})\BibitemShut {NoStop}%
\bibitem [{\citenamefont {De~la Madrid}, \citenamefont {Bohm},\ and\
  \citenamefont {Gadella}(2002)}]{de2002rigged}%
  \BibitemOpen
  \bibfield  {author} {\bibinfo {author} {\bibfnamefont {R.}~\bibnamefont
  {De~la Madrid}}, \bibinfo {author} {\bibfnamefont {A.}~\bibnamefont {Bohm}},\
  and\ \bibinfo {author} {\bibfnamefont {M.}~\bibnamefont {Gadella}},\
  }\bibfield  {title} {\enquote {\bibinfo {title} {Rigged hilbert space
  treatment of continuous spectrum},}\ }\href@noop {} {\bibfield  {journal}
  {\bibinfo  {journal} {Fortschritte der Physik: Progress of Physics}\ }\textbf
  {\bibinfo {volume} {50}},\ \bibinfo {pages} {185--216} (\bibinfo {year}
  {2002})}\BibitemShut {NoStop}%
\bibitem [{\citenamefont {De~la Madrid}(2002)}]{de2002rigged2}%
  \BibitemOpen
  \bibfield  {author} {\bibinfo {author} {\bibfnamefont {R.}~\bibnamefont
  {De~la Madrid}},\ }\bibfield  {title} {\enquote {\bibinfo {title} {Rigged
  hilbert space approach to the schr{\"o}dinger equation},}\ }\href@noop {}
  {\bibfield  {journal} {\bibinfo  {journal} {Journal of Physics A:
  Mathematical and General}\ }\textbf {\bibinfo {volume} {35}},\ \bibinfo
  {pages} {319} (\bibinfo {year} {2002})}\BibitemShut {NoStop}%
\bibitem [{\citenamefont {De~la Madrid}(2003)}]{de2003rigged}%
  \BibitemOpen
  \bibfield  {author} {\bibinfo {author} {\bibfnamefont {R.}~\bibnamefont
  {De~la Madrid}},\ }\bibfield  {title} {\enquote {\bibinfo {title} {The rigged
  hilbert space of the free hamiltonian},}\ }\href@noop {} {\bibfield
  {journal} {\bibinfo  {journal} {International Journal of Theoretical
  Physics}\ }\textbf {\bibinfo {volume} {42}},\ \bibinfo {pages} {2441--2460}
  (\bibinfo {year} {2003})}\BibitemShut {NoStop}%
\bibitem [{\citenamefont {Le{\'o}n}\ \emph {et~al.}(2000)\citenamefont
  {Le{\'o}n}, \citenamefont {Julve}, \citenamefont {Pitanga},\ and\
  \citenamefont {De~Urr{\'\i}es}}]{leon2000time}%
  \BibitemOpen
  \bibfield  {author} {\bibinfo {author} {\bibfnamefont {J.}~\bibnamefont
  {Le{\'o}n}}, \bibinfo {author} {\bibfnamefont {J.}~\bibnamefont {Julve}},
  \bibinfo {author} {\bibfnamefont {P.}~\bibnamefont {Pitanga}},\ and\ \bibinfo
  {author} {\bibfnamefont {F.}~\bibnamefont {De~Urr{\'\i}es}},\ }\bibfield
  {title} {\enquote {\bibinfo {title} {Time of arrival in the presence of
  interactions},}\ }\href@noop {} {\bibfield  {journal} {\bibinfo  {journal}
  {Physical Review A}\ }\textbf {\bibinfo {volume} {61}},\ \bibinfo {pages}
  {062101} (\bibinfo {year} {2000})}\BibitemShut {NoStop}%
\bibitem [{\citenamefont {Peres}(2006)}]{peres2006quantum}%
  \BibitemOpen
  \bibfield  {author} {\bibinfo {author} {\bibfnamefont {A.}~\bibnamefont
  {Peres}},\ }\href@noop {} {\emph {\bibinfo {title} {Quantum theory: concepts
  and methods}}},\ Vol.~\bibinfo {volume} {57}\ (\bibinfo  {publisher}
  {Springer Science \& Business Media},\ \bibinfo {year} {2006})\BibitemShut
  {NoStop}%
\bibitem [{\citenamefont {Gotay}, \citenamefont {Grabowski},\ and\
  \citenamefont {Grundling}(2000)}]{gotay2000obstruction}%
  \BibitemOpen
  \bibfield  {author} {\bibinfo {author} {\bibfnamefont {M.}~\bibnamefont
  {Gotay}}, \bibinfo {author} {\bibfnamefont {J.}~\bibnamefont {Grabowski}},\
  and\ \bibinfo {author} {\bibfnamefont {H.}~\bibnamefont {Grundling}},\
  }\bibfield  {title} {\enquote {\bibinfo {title} {An obstruction to quantizing
  compact symplectic manifolds},}\ }\href@noop {} {\bibfield  {journal}
  {\bibinfo  {journal} {Proceedings of the American Mathematical Society}\
  }\textbf {\bibinfo {volume} {128}},\ \bibinfo {pages} {237--243} (\bibinfo
  {year} {2000})}\BibitemShut {NoStop}%
\bibitem [{\citenamefont {Groenewold}\ and\ \citenamefont
  {Groenewold}(1946)}]{groenewold1946principles}%
  \BibitemOpen
  \bibfield  {author} {\bibinfo {author} {\bibfnamefont {H.~J.}\ \bibnamefont
  {Groenewold}}\ and\ \bibinfo {author} {\bibfnamefont {H.~J.}\ \bibnamefont
  {Groenewold}},\ }\href@noop {} {\emph {\bibinfo {title} {On the principles of
  elementary quantum mechanics}}}\ (\bibinfo  {publisher} {Springer},\ \bibinfo
  {year} {1946})\BibitemShut {NoStop}%
\bibitem [{\citenamefont {Galapon}(2001)}]{Galapon2001}%
  \BibitemOpen
  \bibfield  {author} {\bibinfo {author} {\bibfnamefont {E.~A.}\ \bibnamefont
  {Galapon}},\ }\bibfield  {title} {\enquote {\bibinfo {title}
  {Quantum-classical correspondence of dynamical observables, quantization, and
  the time of arrival correspondence problem},}\ }\href
  {https://doi.org/10.1134/1.1405219} {\bibfield  {journal} {\bibinfo
  {journal} {Optics and Spectroscopy}\ }\textbf {\bibinfo {volume} {91}},\
  \bibinfo {pages} {399--405} (\bibinfo {year} {2001})}\BibitemShut {NoStop}%
\bibitem [{\citenamefont {Galapon}(2004)}]{galapon2004shouldn}%
  \BibitemOpen
  \bibfield  {author} {\bibinfo {author} {\bibfnamefont {E.~A.}\ \bibnamefont
  {Galapon}},\ }\bibfield  {title} {\enquote {\bibinfo {title} {Shouldn’t
  there be an antithesis to quantization?}}\ }\href@noop {} {\bibfield
  {journal} {\bibinfo  {journal} {Journal of mathematical physics}\ }\textbf
  {\bibinfo {volume} {45}},\ \bibinfo {pages} {3180--3215} (\bibinfo {year}
  {2004})}\BibitemShut {NoStop}%
\bibitem [{\citenamefont {Pablico}\ and\ \citenamefont
  {Galapon}(2023)}]{pablico2022quantum}%
  \BibitemOpen
  \bibfield  {author} {\bibinfo {author} {\bibfnamefont {D.~A.~L.}\
  \bibnamefont {Pablico}}\ and\ \bibinfo {author} {\bibfnamefont {E.~A.}\
  \bibnamefont {Galapon}},\ }\bibfield  {title} {\enquote {\bibinfo {title}
  {Quantum corrections to the weyl quantization of the classical time of
  arrival},}\ }\href@noop {} {\bibfield  {journal} {\bibinfo  {journal} {The
  European Physical Journal Plus}\ }\textbf {\bibinfo {volume} {138}},\
  \bibinfo {pages} {1--22} (\bibinfo {year} {2023})}\BibitemShut {NoStop}%
\bibitem [{\citenamefont {Bender}\ and\ \citenamefont
  {Dunne}(1989{\natexlab{a}})}]{bender1989exact}%
  \BibitemOpen
  \bibfield  {author} {\bibinfo {author} {\bibfnamefont {C.~M.}\ \bibnamefont
  {Bender}}\ and\ \bibinfo {author} {\bibfnamefont {G.~V.}\ \bibnamefont
  {Dunne}},\ }\bibfield  {title} {\enquote {\bibinfo {title} {Exact solutions
  to operator differential equations},}\ }\href@noop {} {\bibfield  {journal}
  {\bibinfo  {journal} {Physical Review D}\ }\textbf {\bibinfo {volume} {40}},\
  \bibinfo {pages} {2739} (\bibinfo {year} {1989}{\natexlab{a}})}\BibitemShut
  {NoStop}%
\bibitem [{\citenamefont {Bender}\ and\ \citenamefont
  {Dunne}(1989{\natexlab{b}})}]{bender1989integration}%
  \BibitemOpen
  \bibfield  {author} {\bibinfo {author} {\bibfnamefont {C.~M.}\ \bibnamefont
  {Bender}}\ and\ \bibinfo {author} {\bibfnamefont {G.~V.}\ \bibnamefont
  {Dunne}},\ }\bibfield  {title} {\enquote {\bibinfo {title} {Integration of
  operator differential equations},}\ }\href@noop {} {\bibfield  {journal}
  {\bibinfo  {journal} {Physical Review D}\ }\textbf {\bibinfo {volume} {40}},\
  \bibinfo {pages} {3504} (\bibinfo {year} {1989}{\natexlab{b}})}\BibitemShut
  {NoStop}%
\bibitem [{\citenamefont {Domingo}\ and\ \citenamefont
  {Galapon}(2015)}]{domingo2015generalized}%
  \BibitemOpen
  \bibfield  {author} {\bibinfo {author} {\bibfnamefont {H.~B.}\ \bibnamefont
  {Domingo}}\ and\ \bibinfo {author} {\bibfnamefont {E.~A.}\ \bibnamefont
  {Galapon}},\ }\bibfield  {title} {\enquote {\bibinfo {title} {Generalized
  weyl transform for operator ordering: polynomial functions in phase space},}\
  }\href@noop {} {\bibfield  {journal} {\bibinfo  {journal} {Journal of
  Mathematical Physics}\ }\textbf {\bibinfo {volume} {56}},\ \bibinfo {pages}
  {022104} (\bibinfo {year} {2015})}\BibitemShut {NoStop}%
\bibitem [{\citenamefont {De~Gosson}(2016)}]{de2016born}%
  \BibitemOpen
  \bibfield  {author} {\bibinfo {author} {\bibfnamefont {M.~A.}\ \bibnamefont
  {De~Gosson}},\ }\href@noop {} {\emph {\bibinfo {title} {Born-Jordan
  quantization: theory and applications}}},\ Vol.\ \bibinfo {volume} {182}\
  (\bibinfo  {publisher} {Springer},\ \bibinfo {year} {2016})\BibitemShut
  {NoStop}%
\bibitem [{\citenamefont {Cohen}(2012)}]{cohen2012weyl}%
  \BibitemOpen
  \bibfield  {author} {\bibinfo {author} {\bibfnamefont {L.}~\bibnamefont
  {Cohen}},\ }\href@noop {} {\emph {\bibinfo {title} {The Weyl operator and its
  generalization}}}\ (\bibinfo  {publisher} {Springer Science \& Business
  Media},\ \bibinfo {year} {2012})\BibitemShut {NoStop}%
\bibitem [{\citenamefont {de~Gosson}(2016{\natexlab{a}})}]{Gosson2016}%
  \BibitemOpen
  \bibfield  {author} {\bibinfo {author} {\bibfnamefont {M.~A.}\ \bibnamefont
  {de~Gosson}},\ }\bibfield  {title} {\enquote {\bibinfo {title} {From {{Weyl}}
  to {{Born}}–{{Jordan}} quantization: {{The Schrödinger}} representation
  revisited},}\ }\href@noop {} {\bibfield  {journal} {\bibinfo  {journal}
  {Physics Reports}\ }\textbf {\bibinfo {volume} {623}},\ \bibinfo {pages}
  {1--58} (\bibinfo {year} {2016}{\natexlab{a}})}\BibitemShut {NoStop}%
\bibitem [{\citenamefont {de~Gosson}(2016{\natexlab{b}})}]{Gosson2016a}%
  \BibitemOpen
  \bibfield  {author} {\bibinfo {author} {\bibfnamefont {M.~A.}\ \bibnamefont
  {de~Gosson}},\ }\enquote {\bibinfo {title} {Born–{{Jordan
  Quantization}}},}\ in\ \href@noop {} {\emph {\bibinfo {booktitle}
  {Born-{{Jordan Quantization}}}}},\ \bibinfo {series} {Fundamental
  {{Theories}} of {{Physics}}}, Vol.\ \bibinfo {volume} {182}\ (\bibinfo
  {publisher} {{Springer International Publishing}},\ \bibinfo {address}
  {{Cham}},\ \bibinfo {year} {2016})\ pp.\ \bibinfo {pages}
  {113--127}\BibitemShut {NoStop}%
\bibitem [{\citenamefont {de~Gosson}\ and\ \citenamefont
  {Luef}(2011)}]{Gosson2011}%
  \BibitemOpen
  \bibfield  {author} {\bibinfo {author} {\bibfnamefont {M.}~\bibnamefont
  {de~Gosson}}\ and\ \bibinfo {author} {\bibfnamefont {F.}~\bibnamefont
  {Luef}},\ }\bibfield  {title} {\enquote {\bibinfo {title} {Preferred
  quantization rules: {{Born}}–{{Jordan}} versus {{Weyl}}. {{The}}
  pseudo-differential point of view},}\ }\href@noop {} {\bibfield  {journal}
  {\bibinfo  {journal} {Journal of Pseudo-Differential Operators and
  Applications}\ }\textbf {\bibinfo {volume} {2}},\ \bibinfo {pages} {115--139}
  (\bibinfo {year} {2011})}\BibitemShut {NoStop}%
\bibitem [{\citenamefont {De~Gosson}(2006)}]{de2006symplectic}%
  \BibitemOpen
  \bibfield  {author} {\bibinfo {author} {\bibfnamefont {M.~A.}\ \bibnamefont
  {De~Gosson}},\ }\href@noop {} {\emph {\bibinfo {title} {Symplectic geometry
  and quantum mechanics}}},\ Vol.\ \bibinfo {volume} {166}\ (\bibinfo
  {publisher} {Springer Science \& Business Media},\ \bibinfo {year}
  {2006})\BibitemShut {NoStop}%
\bibitem [{\citenamefont {De~Gosson}(2013)}]{de2013born}%
  \BibitemOpen
  \bibfield  {author} {\bibinfo {author} {\bibfnamefont {M.~A.}\ \bibnamefont
  {De~Gosson}},\ }\bibfield  {title} {\enquote {\bibinfo {title} {Born--jordan
  quantization and the uncertainty principle},}\ }\href@noop {} {\bibfield
  {journal} {\bibinfo  {journal} {Journal of Physics A: Mathematical and
  Theoretical}\ }\textbf {\bibinfo {volume} {46}},\ \bibinfo {pages} {445301}
  (\bibinfo {year} {2013})}\BibitemShut {NoStop}%
\bibitem [{\citenamefont {Cohen}(1966)}]{cohen1966generalized}%
  \BibitemOpen
  \bibfield  {author} {\bibinfo {author} {\bibfnamefont {L.}~\bibnamefont
  {Cohen}},\ }\bibfield  {title} {\enquote {\bibinfo {title} {Generalized
  phase-space distribution functions},}\ }\href@noop {} {\bibfield  {journal}
  {\bibinfo  {journal} {Journal of Mathematical Physics}\ }\textbf {\bibinfo
  {volume} {7}},\ \bibinfo {pages} {781--786} (\bibinfo {year}
  {1966})}\BibitemShut {NoStop}%
\bibitem [{\citenamefont {Shewell}(1959)}]{shewell1959formation}%
  \BibitemOpen
  \bibfield  {author} {\bibinfo {author} {\bibfnamefont {J.~R.}\ \bibnamefont
  {Shewell}},\ }\bibfield  {title} {\enquote {\bibinfo {title} {On the
  formation of quantum-mechanical operators},}\ }\href@noop {} {\bibfield
  {journal} {\bibinfo  {journal} {American Journal of Physics}\ }\textbf
  {\bibinfo {volume} {27}},\ \bibinfo {pages} {16--21} (\bibinfo {year}
  {1959})}\BibitemShut {NoStop}%
\bibitem [{\citenamefont {Greiner}\ \emph {et~al.}(2000)\citenamefont {Greiner}
  \emph {et~al.}}]{greiner2000relativistic}%
  \BibitemOpen
  \bibfield  {author} {\bibinfo {author} {\bibfnamefont {W.}~\bibnamefont
  {Greiner}} \emph {et~al.},\ }\href@noop {} {\emph {\bibinfo {title}
  {Relativistic quantum mechanics}}},\ Vol.~\bibinfo {volume} {2}\ (\bibinfo
  {publisher} {Springer},\ \bibinfo {year} {2000})\BibitemShut {NoStop}%
\bibitem [{\citenamefont {Le{\'o}n}(1997)}]{leon1997time}%
  \BibitemOpen
  \bibfield  {author} {\bibinfo {author} {\bibfnamefont {J.}~\bibnamefont
  {Le{\'o}n}},\ }\bibfield  {title} {\enquote {\bibinfo {title}
  {Time-of-arrival formalism for the relativistic particle},}\ }\href@noop {}
  {\bibfield  {journal} {\bibinfo  {journal} {Journal of Physics A:
  Mathematical and General}\ }\textbf {\bibinfo {volume} {30}},\ \bibinfo
  {pages} {4791} (\bibinfo {year} {1997})}\BibitemShut {NoStop}%
\bibitem [{\citenamefont {Newton}\ and\ \citenamefont
  {Wigner}(1949)}]{newton1949localized}%
  \BibitemOpen
  \bibfield  {author} {\bibinfo {author} {\bibfnamefont {T.~D.}\ \bibnamefont
  {Newton}}\ and\ \bibinfo {author} {\bibfnamefont {E.~P.}\ \bibnamefont
  {Wigner}},\ }\bibfield  {title} {\enquote {\bibinfo {title} {Localized states
  for elementary systems},}\ }\href@noop {} {\bibfield  {journal} {\bibinfo
  {journal} {Reviews of Modern Physics}\ }\textbf {\bibinfo {volume} {21}},\
  \bibinfo {pages} {400} (\bibinfo {year} {1949})}\BibitemShut {NoStop}%
\bibitem [{\citenamefont {Razavy}(1969)}]{razavy1969quantum}%
  \BibitemOpen
  \bibfield  {author} {\bibinfo {author} {\bibfnamefont {M.}~\bibnamefont
  {Razavy}},\ }\bibfield  {title} {\enquote {\bibinfo {title}
  {Quantum-mechanical conjugate of the hamiltonian operator},}\ }\href@noop {}
  {\bibfield  {journal} {\bibinfo  {journal} {Il Nuovo Cimento B (1965-1970)}\
  }\textbf {\bibinfo {volume} {63}},\ \bibinfo {pages} {271--308} (\bibinfo
  {year} {1969})}\BibitemShut {NoStop}%
\bibitem [{\citenamefont {Flores}\ and\ \citenamefont
  {Galapon}(2022{\natexlab{a}})}]{flores2022relativistic}%
  \BibitemOpen
  \bibfield  {author} {\bibinfo {author} {\bibfnamefont {P.~C.}\ \bibnamefont
  {Flores}}\ and\ \bibinfo {author} {\bibfnamefont {E.~A.}\ \bibnamefont
  {Galapon}},\ }\bibfield  {title} {\enquote {\bibinfo {title} {Relativistic
  free-motion time-of-arrival operator for massive spin-0 particles with
  positive energy},}\ }\href@noop {} {\bibfield  {journal} {\bibinfo  {journal}
  {Physical Review A}\ }\textbf {\bibinfo {volume} {105}},\ \bibinfo {pages}
  {062208} (\bibinfo {year} {2022}{\natexlab{a}})}\BibitemShut {NoStop}%
\bibitem [{\citenamefont {Galapon}(2016)}]{Galapon2016}%
  \BibitemOpen
  \bibfield  {author} {\bibinfo {author} {\bibfnamefont {E.~A.}\ \bibnamefont
  {Galapon}},\ }\bibfield  {title} {\enquote {\bibinfo {title} {The {{Cauchy}}
  principal value and the {{Hadamard}} finite part integral as values of
  absolutely convergent integrals},}\ }\href@noop {} {\bibfield  {journal}
  {\bibinfo  {journal} {Journal of Mathematical Physics}\ }\textbf {\bibinfo
  {volume} {57}},\ \bibinfo {pages} {033502} (\bibinfo {year}
  {2016})}\BibitemShut {NoStop}%
\bibitem [{\citenamefont {Sombillo}\ and\ \citenamefont
  {Galapon}(2014)}]{sombillo2014quantum}%
  \BibitemOpen
  \bibfield  {author} {\bibinfo {author} {\bibfnamefont {D.~L.}\ \bibnamefont
  {Sombillo}}\ and\ \bibinfo {author} {\bibfnamefont {E.~A.}\ \bibnamefont
  {Galapon}},\ }\bibfield  {title} {\enquote {\bibinfo {title} {Quantum
  traversal time through a double barrier},}\ }\href@noop {} {\bibfield
  {journal} {\bibinfo  {journal} {Physical Review A}\ }\textbf {\bibinfo
  {volume} {90}},\ \bibinfo {pages} {032115} (\bibinfo {year}
  {2014})}\BibitemShut {NoStop}%
\bibitem [{\citenamefont {Sombillo}\ and\ \citenamefont
  {Galapon}(2018)}]{sombillo2018barrier}%
  \BibitemOpen
  \bibfield  {author} {\bibinfo {author} {\bibfnamefont {D.~L.~B.}\
  \bibnamefont {Sombillo}}\ and\ \bibinfo {author} {\bibfnamefont {E.~A.}\
  \bibnamefont {Galapon}},\ }\bibfield  {title} {\enquote {\bibinfo {title}
  {Barrier-traversal-time operator and the time-energy uncertainty relation},}\
  }\href@noop {} {\bibfield  {journal} {\bibinfo  {journal} {Physical Review
  A}\ }\textbf {\bibinfo {volume} {97}},\ \bibinfo {pages} {062127} (\bibinfo
  {year} {2018})}\BibitemShut {NoStop}%
\bibitem [{\citenamefont {Gel'fand}\ and\ \citenamefont
  {Shi}(1964)}]{gel1964ov}%
  \BibitemOpen
  \bibfield  {author} {\bibinfo {author} {\bibfnamefont {I.}~\bibnamefont
  {Gel'fand}}\ and\ \bibinfo {author} {\bibfnamefont {G.}~\bibnamefont {Shi}},\
  }\enquote {\bibinfo {title} {ov, generalized functions. vol. i: Properties
  and operations},}\ \ (\bibinfo  {publisher} {Academic Press, New York},\
  \bibinfo {year} {1964})\ p.\ \bibinfo {pages} {360}\BibitemShut {NoStop}%
\bibitem [{\citenamefont {Flores}\ and\ \citenamefont
  {Galapon}(2022{\natexlab{b}})}]{PhysRevA.105.062208}%
  \BibitemOpen
  \bibfield  {author} {\bibinfo {author} {\bibfnamefont {P.~C.}\ \bibnamefont
  {Flores}}\ and\ \bibinfo {author} {\bibfnamefont {E.~A.}\ \bibnamefont
  {Galapon}},\ }\bibfield  {title} {\enquote {\bibinfo {title} {Relativistic
  free-motion time-of-arrival operator for massive spin-0 particles with
  positive energy},}\ }\href {https://doi.org/10.1103/PhysRevA.105.062208}
  {\bibfield  {journal} {\bibinfo  {journal} {Phys. Rev. A}\ }\textbf {\bibinfo
  {volume} {105}},\ \bibinfo {pages} {062208} (\bibinfo {year}
  {2022}{\natexlab{b}})}\BibitemShut {NoStop}%
\bibitem [{\citenamefont {Pablico}\ and\ \citenamefont
  {Galapon}(2020)}]{pablico2020quantum}%
  \BibitemOpen
  \bibfield  {author} {\bibinfo {author} {\bibfnamefont {D.~A.~L.}\
  \bibnamefont {Pablico}}\ and\ \bibinfo {author} {\bibfnamefont {E.~A.}\
  \bibnamefont {Galapon}},\ }\bibfield  {title} {\enquote {\bibinfo {title}
  {Quantum traversal time across a potential well},}\ }\href@noop {} {\bibfield
   {journal} {\bibinfo  {journal} {Physical Review A}\ }\textbf {\bibinfo
  {volume} {101}},\ \bibinfo {pages} {022103} (\bibinfo {year}
  {2020})}\BibitemShut {NoStop}%
\bibitem [{\citenamefont {Galapon}(2017)}]{galapon2017problem}%
  \BibitemOpen
  \bibfield  {author} {\bibinfo {author} {\bibfnamefont {E.~A.}\ \bibnamefont
  {Galapon}},\ }\bibfield  {title} {\enquote {\bibinfo {title} {The problem of
  missing terms in term by term integration involving divergent integrals},}\
  }\href@noop {} {\bibfield  {journal} {\bibinfo  {journal} {Proceedings of the
  Royal Society A: Mathematical, Physical and Engineering Sciences}\ }\textbf
  {\bibinfo {volume} {473}},\ \bibinfo {pages} {20160567} (\bibinfo {year}
  {2017})}\BibitemShut {NoStop}%
\bibitem [{\citenamefont {Tica}\ and\ \citenamefont
  {Galapon}(2019)}]{tica2019finite}%
  \BibitemOpen
  \bibfield  {author} {\bibinfo {author} {\bibfnamefont {C.~D.}\ \bibnamefont
  {Tica}}\ and\ \bibinfo {author} {\bibfnamefont {E.~A.}\ \bibnamefont
  {Galapon}},\ }\bibfield  {title} {\enquote {\bibinfo {title} {Finite-part
  integration of the generalized stieltjes transform and its dominant
  asymptotic behavior for small values of the parameter. ii. non-integer
  orders},}\ }\href@noop {} {\bibfield  {journal} {\bibinfo  {journal} {Journal
  of Mathematical Physics}\ }\textbf {\bibinfo {volume} {60}},\ \bibinfo
  {pages} {013502} (\bibinfo {year} {2019})}\BibitemShut {NoStop}%
\bibitem [{\citenamefont {Bunao}\ and\ \citenamefont
  {Galapon}(2015{\natexlab{a}})}]{bunao2015one}%
  \BibitemOpen
  \bibfield  {author} {\bibinfo {author} {\bibfnamefont {J.}~\bibnamefont
  {Bunao}}\ and\ \bibinfo {author} {\bibfnamefont {E.~A.}\ \bibnamefont
  {Galapon}},\ }\bibfield  {title} {\enquote {\bibinfo {title} {A one-particle
  time of arrival operator for a free relativistic spin-0 charged particle in
  (1+ 1) dimensions},}\ }\href@noop {} {\bibfield  {journal} {\bibinfo
  {journal} {Annals of Physics}\ }\textbf {\bibinfo {volume} {353}},\ \bibinfo
  {pages} {83--106} (\bibinfo {year} {2015}{\natexlab{a}})}\BibitemShut
  {NoStop}%
\bibitem [{\citenamefont {Bunao}\ and\ \citenamefont
  {Galapon}(2015{\natexlab{b}})}]{bunao2015relativistic}%
  \BibitemOpen
  \bibfield  {author} {\bibinfo {author} {\bibfnamefont {J.}~\bibnamefont
  {Bunao}}\ and\ \bibinfo {author} {\bibfnamefont {E.~A.}\ \bibnamefont
  {Galapon}},\ }\bibfield  {title} {\enquote {\bibinfo {title} {A relativistic
  one-particle time of arrival operator for a free spin-1/2 particle in (1+ 1)
  dimensions},}\ }\href@noop {} {\bibfield  {journal} {\bibinfo  {journal}
  {Annals of Physics}\ }\textbf {\bibinfo {volume} {356}},\ \bibinfo {pages}
  {369--382} (\bibinfo {year} {2015}{\natexlab{b}})}\BibitemShut {NoStop}%
\bibitem [{\citenamefont {Farrales}, \citenamefont {Domingo},\ and\
  \citenamefont {Galapon}(2022)}]{farrales2022conjugates}%
  \BibitemOpen
  \bibfield  {author} {\bibinfo {author} {\bibfnamefont {R.~A.~E.}\
  \bibnamefont {Farrales}}, \bibinfo {author} {\bibfnamefont {H.~B.}\
  \bibnamefont {Domingo}},\ and\ \bibinfo {author} {\bibfnamefont {E.~A.}\
  \bibnamefont {Galapon}},\ }\bibfield  {title} {\enquote {\bibinfo {title}
  {Conjugates to one particle hamiltonians in 1-dimension in differential
  form},}\ }\href@noop {} {\bibfield  {journal} {\bibinfo  {journal} {The
  European Physical Journal Plus}\ }\textbf {\bibinfo {volume} {137}},\
  \bibinfo {pages} {1--24} (\bibinfo {year} {2022})}\BibitemShut {NoStop}%
\end{thebibliography}%

\end{document}